\documentclass[12pt]{article}
\pdfoutput=1

\usepackage{draft} 
\usepackage{hyperref}
\usepackage{graphicx,color,subfig}
\usepackage{cite}
\usepackage{mciteplus}
\usepackage{skak}
\usepackage{empheq}
\usepackage{anyfontsize}
\usepackage{amsthm}
\newtheorem{theo}{Theorem}
\newtheorem{define}{Definition}
\newtheorem{case}{Case}

\DeclareFontFamily{OT1}{pzc}{}
\DeclareFontShape{OT1}{pzc}{m}{it}{<-> s * [1.10] pzcmi7t}{}
\DeclareMathAlphabet{\mathpzc}{OT1}{pzc}{m}{it}

\def\be#1\ee{\begin{align}#1\end{align}}

\begin{document}

\unitlength = .8mm

\begin{titlepage}

\begin{center}

\hfill \\
\hfill \\
\vskip 1cm

\title{Open-closed Hyperbolic String Vertices}

\author{Minjae Cho}

\address{
Jefferson Physical Laboratory, Harvard University, \\
Cambridge, MA 02138 USA
}

\email{minjaecho@fas.harvard.edu}

\end{center}

\abstract{We construct a family of hyperbolic string vertices in the oriented open-closed string field theory, generalizing the recent result on hyperbolic closed string vertices by Costello and Zwiebach. The vertices are described by certain bordered hyperbolic surfaces and we explain relevant collar theorems which provide restrictions on the systolic conditions for the hyperbolic vertices. We also give explicit descriptions of the vertices for all zero and one-dimensional moduli spaces.
}

\vfill

\end{titlepage}

\eject

\begingroup
\hypersetup{linkcolor=black}

\tableofcontents

\endgroup

\section{Introduction} 
In recent years, it has become clear that a well defined and consistent perturbative formulation of string theory requires the framework of string field theory (we refer readers to \cite{Zwiebach:1992ie, Sen:2015uaa,deLacroix:2017lif} and references therein for an overview of the subject). For example, traditional analytic continuation involved in the computation of string amplitudes to cure the divergences in moduli integration naturally arises in string field theory as explained in \cite{Sen:2019jpm}. Furthermore, prescriptions given by string field theory provide not only unambiguous recipe to compute physical quantities in a given background, but also descriptions of more general backgrounds arising as solutions to string field equations of motions. For example, this idea was used to study strings in Ramond-Ramond flux backgrounds \cite{Cho:2018nfn}.

In practice, computations that appear in string field theory are those of worldsheet conformal field theories. Therefore, we in principle can compute relevant quantities in a rather strightforward manner. Being a field theory, string field theory carries vertices which are roughly speaking integration of worldsheet correlators of off-shell string fields over specific parts of the moduli spaces. However, such off-shell objects in general depend on how one coordinatizes Riemann surfaces and this ambiguity exactly amounts to string field redefinitions \cite{Hata:1993gf,Sen:1993ic,Sen:2014dqa,Sen:2015hha}. Thus, the choice of vertices amounts to which coordinatization of Riemann surfaces to use and which parts of the moduli spaces to cover.

Of course, not all such arbitrary choices of vertices are consistent. There is a very natural requirement on string vertices when the homomorphism between Batalin-Vilkoviski (BV)-algebras of surfaces and string fields are considered \cite{Sen:1994kx,Sen:1993kb}. The requirement is called geometric master equation and the job of finding the solutions is of fundamental interest in the framework of string field theory. In the past, such solutions were found using various metrics, an example being minimal area metrics \cite{Zwiebach:1990nh,Zwiebach:1990qj,Zwiebach:1992bw,Headrick:2018ncs,Headrick:2018dlw}. There were also approximate constructions using the hyperbolic metrics \cite{Moosavian:2017fta,Moosavian:2017qsp,Moosavian:2017sev,Pius:2018pqr}.

Recently, a nice explicit construction of closed string vertices using hyperbolic metric was achieved in \cite{Costello:2019fuh}. One starts with a bordered hyperbolic Riemann surface with specified border lengths and systolic constraints, and grafts flat semi-infinite cylinders to the borders to make them into punctures. Upon connecting such vertices using closed string propagator which is represented as a flat finite cylinder, the resulting metric is the Thurston metric (for an overview, see \cite{tanigawa1995grafting}). As string theory requires us to integrate over the moduli space, one possible advantage of hyperbolic vertices from string theory perspective is that there is a better understanding of moduli integration in such metrics \cite{1998InMat.132..607M,Mirzakhani:2006fta,Mirzakhani:2006eta}.

In this work, we generalize the construction of hyperbolic string vertices to oriented open-closed string field theory \cite{Zwiebach:1990qj,Zwiebach:1997fe} (we will omit the term "oriented" from now on and it is always assumed). Hyperbolic surfaces to be considered are bordered hyperbolic surfaces (we refer readers to Chapter 1 of \cite{Buser1992GeometryAS} for a gentle introduction to these surfaces). Their boundaries are piecewise geodesic and some of geodesic sides will correspond to open string punctures, while the other sides belong to boundaries. Some of the borders which are closed geodesics will correspond to closed string punctures as in \cite{Costello:2019fuh}, while the others will correspond to boundaries. We will define a family of subsets of such bordered hyperbolic surfaces, and show that it solves the geometric master equation. The essential ingredients of the proof are collar theorems. Such theorems are well-known for hyperbolic bordered Riemann surfaces where boundaries are all smooth closed geodesics. We will extend them to the case of bordered hyperbolic surfaces under consideration. We will also give explicit description of all zero and one-dimensional open-closed hyperbolic string vertices. This description will show that the family of hyperbolic vertices we constructed do not include a point where the theory becomes Witten's cubic theory \cite{Witten:1985cc,Zwiebach:1992bw}, as already discussed in \cite{Costello:2019fuh}.

The paper is organized as follows. In section \ref{geometricmastereq}, we review the geometric master equation for open-closed string vertices \cite{Zwiebach:1997fe} and provide a proof that Feynman diagrams built out of the solutions to the geometric master equation represent fundamental classes in relative homologies of interest. Then, we proceed to discuss relevant geometric objects and theorems in section \ref{hyperbolicsurfaces}. Using these objects, in section \ref{hyperbolicvertices} we will define hyperbolic open-closed string vertices and prove that they solve the geometric master equation. We describe zero and one-dimensional hyperbolic vertices in section \ref{lowdimvertices}. We conclude with remarks and discussions in section \ref{discussions}.

\section{Open-closed geometric master equation and Feynman diagrams}\label{geometricmastereq}
In this section, we review the general framework of open-closed string field theory and the corresponding geometric master equation, which we will solve in later sections. All discussions are standard and we will closely follow \cite{Zwiebach:1997fe}. Then, we will show that the Feynman diagrams built out of the solutions cover the moduli space exactly once, following the ideas presented in \cite{Costello:2019fuh}.

\subsection{Moduli spaces, total spaces, and singular chains}
In open-closed string field theory, the geometric objects under consideration are bordered Riemann surfaces with marked bulk and boundary punctures. In order to specify the moduli space, one specifies genus $g$, number of bulk punctures $n$, number of boundary components $b$, and number $m_i$ of boundary punctures on the $i$-th boundary, with $i=1,2,...,b$. We will denote the corresponding moduli space as ${\cal M}^{g,n}_{b,\{m_i\}}$. Open-closed string vertices take all possible values of these parameters satisfying
\ie\label{moduliconditions}
&i)~n\geq3~\text{for}~g=b=0,
\\
&ii)~n\geq1~\text{for}~g=1,~b=0,
\\
&iii)~m_1\geq3~\text{for}~g=0,~b=1.
\fe

Off-shell amplitudes of string fields are integration of worldsheet correlators over a given moduli space. As already mentioned, the result depends on the coordinatization of Riemann surfaces. We first introduce local coordinates around bulk and boundary punctures. For bulk punctures, we will take flat unit disk $\{z\in\mathbb{C}|~|z|\leq1\}$ whose origin is the location of the puncture. For boundary punctures, we will take semi-disk $\{z\in\mathbb{C}|~|z|\leq1~\text{and}~\text{Im}(z)\geq0\}$ on flat upper-half plane, where the origin is the location of the puncture and the real axis is the boundary. Then, the choice of embedding of disks and semi-disks corresponds to coordinatization of Riemann surfaces. Over the moduli space ${\cal M}^{g,n}_{b,\{m_i\}}$, we will take the fiber to be such embeddings modulo phase rotations for disk coordinates around bulk punctures. The resulting total space is denoted as $\hat{\cal P}^{g,n}_{b,\{m_i\}}$. By forgetting about coordinates, one can naturally project down to the moduli space, $\pi:\hat{\cal P}^{g,n}_{b,\{m_i\}}\rightarrow{\cal M}^{g,n}_{b,\{m_i\}}$.

Typically, one would choose a section in $\hat{\cal P}^{g,n}_{b,\{m_i\}}$ over ${\cal M}^{g,n}_{b,\{m_i\}}$ and compute off-shell amplitudes by integrating along the section. However, as pointed out in \cite{Costello:2019fuh}, one in general can allow for singular chains with real coefficients, as chains are natural objects to integrate over. We also assume that chains are symmetrized over the punctures.

\subsection{Vertices and Feynman diagrams}
Say we made a choice of chains for all zero dimensional moduli spaces, where chains are in the fundamental homology class of the corresponding moduli spaces when pushed forward to it. Roughly speaking, it means that chains cover the moduli space (which is a point here) exactly once taking into account the multiplicities. Then, we can construct Feynman diagrams by combining these vertices using either open string or closed string propagators. Explicitly, closed string propagator plumbs two bulk punctures with local disk coordinates $z$ and $w$ via $zw=e^{-s+i\theta}$ for all $s\geq0$ and $0<\theta\leq2\pi$, and open string propagator glues two boundary punctures with local semi-disk coordinates $z$ and $w$ via $zw=-e^{-s}$ for all $s\geq0$. As a result, such Feynman diagrams are equipped with specific coordinate systems and thus represent chains over higher dimensional moduli spaces. These chains, when pushed forward to the moduli space, in general do not belong to the relative homology $H_{\text{dim}({\cal M})}\left(\overline{\cal M}; \partial\overline{\cal M}\right)$, where $\overline{\cal M}$ is the Deligne-Mumford compactification \cite{Mumford1983TowardsAE} of the moduli space ${\cal M}$ under consideration. Roughly speaking, the moduli space is not covered with multiplicity one by these Feynman diagrams. The only known exception is Witten's cubic theory \cite{Witten:1985cc} where Feynman diagrams built out of propagators and cubic open string vertex cover the entire moduli spaces of all disk diagrams with boundary punctures exactly once. Therefore, one needs to further add chains, which we call vertices, so that the sum of vertices and Feynman diagrams with propagators represents fundamental classes of the homology relative to the boundary, when pushed forward to the moduli space (there is an ambiguity in the notion of the fundamental class of the relative homology which we will properly discuss in section \ref{fundclass}). The generalization to general ${\cal M}^{g,n}_{b,\{m_i\}}$ is straightforward.

In the discussion so far, there is no reason why vertex and Feynman diagrams with propagators should be disjoint and continuous across where they meet. Indeed, they may even share some of the regions over the moduli space. However, BV algebra defined over such chains is naturally homomorphic to that over string fields \cite{Sen:1994kx,Sen:1993kb}. Thus, one requires that the homomorphic preimage of BV master equation for string fields to hold true for chains under consideration here. This is exactly the geometric master equation which we discuss now.

\subsection{Geometric master equation}
In open-closed string field theory, the operations appearing in the BV algebra of chains are $\{{\cal V}_1,{\cal V}_2\}_o,\{{\cal V}_1,{\cal V}_2\}_c,\Delta_o{\cal V},$ and $\Delta_c{\cal V}$ for chains ${\cal V},{\cal V}_1,{\cal V}_2$. For detailed definition and discussions of these operations, see \cite{Sen:1994kx,Zwiebach:1997fe}. We will give a brief description of these operations.

$\{{\cal V}_1,{\cal V}_2\}_o$ glues an open string puncture in every element of the chain ${\cal V}_1$ to another open string puncture in every element of the chain ${\cal V}_2$ using the semi-disk local coordinate identification $zw=-1$. The result will be another chain of dimension same as sum of dimensions of ${\cal V}_1$ and ${\cal V}_2$.

$\{{\cal V}_1,{\cal V}_2\}_c$ twist-plumbs a closed string puncture in every element of the chain ${\cal V}_1$ to another closed string puncture in every element of the chain ${\cal V}_2$ using the disk local coordinate identification $zw=e^{i\theta}$ for all $0<\theta\leq2\pi$. The result will be another chain of dimension higher than the sum of dimensions of ${\cal V}_1$ and ${\cal V}_2$ by one, except for the case where one of the chains, say ${\cal V}_1$, is over ${\cal M}^{0,1}_{1,\{0\}}$, i.e. disk with a bulk puncture, whose moduli space is of dimension zero and has a conformal Killing vector. In the latter case, dimension of the resulting chain is the same as that of ${\cal V}_2$.

$\Delta_o{\cal V}$ glues an open string puncture in every element of the chain ${\cal V}$ to another open string puncture in the same element of the same chain ${\cal V}$ using the semi-disk local coordinate identification $zw=-1$. The result will be another chain of dimension same as that of ${\cal V}$.

$\Delta_c{\cal V}$ twist-plumbs a closed string puncture in every element of the chain ${\cal V}$ to another closed string puncture in the same element of the same chain ${\cal V}$ using the disk local coordinate identification $zw=e^{i\theta}$ for all $0<\theta\leq2\pi$. The result will be another chain of dimension higher than that of ${\cal V}$ by one.

For the convenience of notations, we introduce $\{~,~\}\equiv\{~,~\}_o+\{~,~\}_c$ and $\Delta\equiv\Delta_o+\Delta_c$. Also, there is a natural boundary operator acting on the chain in the homological sense, which we denote as $\partial\cal V$. Now, we introduce the following formal sum of chains over all moduli spaces obeying (\ref{moduliconditions})
\ie
{\cal V}=\sum_{g,n,b,\{m_i\}}\hbar^p\kappa^q{\cal V}^{g,n}_{b,\{m_i\}},
\fe
where $\kappa$ is the string coupling and ${\cal V}^{g,n}_{b,\{m_i\}}$ is a chain in $\hat{\cal P}^{g,n}_{b,\{m_i\}}$ whose dimension is the same as that of ${\cal M}^{g,n}_{b,\{m_i\}}$. The powers $p$ and $q$ are given by $p=2g+{n\over2}+b-1$ and $q=4g+2n+2b+\sum_im_i-4$. Then, the geometric master equation reads
\ie\label{geometricmaster}
\partial{\cal V}+{1\over2}\{{\cal V},{\cal V}\}+\hbar\Delta{\cal V}=0.
\fe
The solution $\cal V$ to (\ref{geometricmaster}) is called string vertices. Geometric master equation (\ref{geometricmaster}) is crucial in deriving Ward identities for off-shell amplitudes, from which null-state decoupling from on-shell amplitudes can be deduced. Therefore, it is tied to the gauge invariance of the string field action.

In case where $\cal V$ are sections solving (\ref{geometricmaster}) and Feynman diagrams built by $\cal V$ are also sections, this condition is a matching condition at the boundary of vertices and Feynman diagrams with propagators. For general chains, the Feynman diagrams constructed by the solutions will represent fundamental classes in the homology relative to the boundary as we discuss now.

\subsection{Fundamental class in homology relative to the boundary}\label{fundclass}
In this subsection, we discuss how Feynman diagrams built using the solution to the geometric master equation (\ref{geometricmaster}) represent fundamental classes in the homology relative to the boundary. It roughly means that they cover the moduli space exactly once. The idea here closely follows similar ideas in \cite{Costello:2019fuh}, but there will be differences arising from infinite length open string propagators and also infinite length closed string propagator connecting to a disk with a bulk puncture. Feynman diagrams $F_{b,\{m_i\}}^{g,n}$ built using vertices are chains in the total space whose dimension is the same as the base moduli space: $F_{b,\{m_i\}}^{g,n}\in C_{\text{dim}({\cal M}_{b,\{m_i\}}^{g,n})}(\hat{\cal P}_{b,\{m_i\}}^{g,n})$. Now, we further include to the chain infinite cylinder and infinite strip propagators. The resulting chain $\overline{F}_{b,\{m_i\}}^{g,n}$ is a chain in the compactified total space $\hat{\overline{{\cal P}}}_{b,\{m_i\}}^{g,n}$, which is a fibered space over the Deligne-Mumford compactification $\overline{\cal M}_{b,\{m_i\}}^{g,n}$ of the moduli space with fibers being the coordinates around bulk and boundary punctures modulo rotation around bulk punctures: $\overline{F}_{b,\{m_i\}}^{g,n}\in C_{\text{dim}({\cal M}_{b,\{m_i\}}^{g,n})}(\hat{\overline{\cal P}}_{b,\{m_i\}}^{g,n})$.

Projection map from the total space to the moduli space naturally extends to the compactified ones, $\pi:\hat{\overline{\cal P}}_{b,\{m_i\}}^{g,n}\rightarrow \overline{\cal M}_{b,\{m_i\}}^{g,n}$. Using this, we push forward $\overline{F}_{b,\{m_i\}}^{g,n}$ to the Delign-Mumford compactified moduli space: $\pi_*\overline{F}_{b,\{m_i\}}^{g,n}\in C_{\text{dim}({\cal M}_{b,\{m_i\}}^{g,n})}(\overline{\cal M}_{b,\{m_i\}}^{g,n})$. We now show that this chain belongs to the homology relative to the boundary $H_{\text{dim}({\cal M}_{b,\{m_i\}}^{g,n})}\left(\overline{\cal M}_{b,\{m_i\}}^{g,n}; \partial\overline{\cal M}_{b,\{m_i\}}^{g,n}\right)$, or in other words, $\partial\left(\pi_*\overline{F}_{b,\{m_i\}}^{g,n}\right)\in C_{\text{dim}({\cal M}_{b,\{m_i\}}^{g,n})-1}\left(\partial\overline{\cal M}_{b,\{m_i\}}^{g,n}\right)$.

Boundaries of the chains can come from three sources: boundaries of vertices, a propagator of length zero, and a propagator of length infinity. The first two cancel each other due to the geometric master equation (\ref{geometricmaster}), and the last belongs to $C_{\text{dim}({\cal M}_{b,\{m_i\}}^{g,n})-1}\left(\partial\overline{\cal M}_{b,\{m_i\}}^{g,n}\right)$. Therefore, the chain under consideration indeed is a cycle in the relative homology. The question now is if it coincides with the fundamental class of the relative homology.

However, the relative homology $H_{\text{dim}({\cal M}_{b,\{m_i\}}^{g,n})}\left(\overline{\cal M}_{b,\{m_i\}}^{g,n}; \partial\overline{\cal M}_{b,\{m_i\}}^{g,n}\right)$ is a product of $\mathbb{R}$ rather than a single $\mathbb{R}$, because the moduli space ${\cal M}_{b,\{m_i\}}^{g,n}$ is disconnected generically. For example, the familiar disk four point function computation of the Veneziano amplitude involves summing over six disks, which correspond to marking inequivalent connected pieces of the disconnected moduli space ${\cal M}_{1,\{4\}}^{0,0}$. Nonetheless, there is a very natural definition of the fundamental class we can take. We denote each connected component of the moduli space as  ${}_I{\cal M}_{b,\{m_i\}}^{g,n}$, where $I$ is an element of an appropriate index set, and consider its Deligne-Mumford compactification ${}_I\overline{\cal M}_{b,\{m_i\}}^{g,n}$. Then $H_{\text{dim}({\cal M}_{b,\{m_i\}}^{g,n})}\left({}_I\overline{\cal M}_{b,\{m_i\}}^{g,n}; \partial{}_I\overline{\cal M}_{b,\{m_i\}}^{g,n}\right)=\mathbb{R}$, since ${}_I\overline{\cal M}_{b,\{m_i\}}^{g,n}$ is a compact connected orbifold. Therefore, in each connected component, there is a unique fundamental class in the homology relative to the boundary. We simply define the fundamental class of $H_{\text{dim}({\cal M}_{b,\{m_i\}}^{g,n})}\left(\overline{\cal M}_{b,\{m_i\}}^{g,n}; \partial\overline{\cal M}_{b,\{m_i\}}^{g,n}\right)$ to be the unique cycle which is the sum of fundamental classes of each connected component of the moduli space with coefficents being 1. This notion of fundamental class is what we want, as this is the precise meaning of covering the moduli space exactly once. Thus, our goal is to show that the cycle $[\pi_*\overline{F}_{b,\{m_i\}}^{g,n}]\in H_{\text{dim}({\cal M}_{b,\{m_i\}}^{g,n})}\left(\overline{\cal M}_{b,\{m_i\}}^{g,n}; \partial\overline{\cal M}_{b,\{m_i\}}^{g,n}\right)$ is the same as the fundamental class we defined.

In order to show that, we should first disconnect the cycle. Over each ${}_I\overline{\cal M}_{b,\{m_i\}}^{g,n}$, the chain $\overline{F}_{b,\{m_i\}}^{g,n}$ will have corresponding part, which we denote ${}_I\overline{F}_{b,\{m_i\}}^{g,n}$, obtained by the restriction map. By similar arguments as before, we have $[\pi_*{}_I\overline{F}_{b,\{m_i\}}^{g,n}]\in H_{\text{dim}({\cal M}_{b,\{m_i\}}^{g,n})}\left({}_I\overline{\cal M}_{b,\{m_i\}}^{g,n}; \partial{}_I\overline{\cal M}_{b,\{m_i\}}^{g,n}\right)=\mathbb{R}$. To show that the multiplicative constant here is 1, it is enough to check one specific point on the chain $[\pi_*{}_I\overline{F}_{b,\{m_i\}}^{g,n}]$. We go back to the entire chain $\overline{F}_{b,\{m_i\}}^{g,n}$. There is a point on the chain over $\overline{\cal M}_{b,\{m_i\}}^{g,n}-{\cal M}_{b,\{m_i\}}^{g,n}$ where the surface is totally degenerate in that it is built by gluing zero-dimensional vertices using infinite length open and closed string propagators. We can consider its neighborhood whose coordinates are the lengths and twists (the latter applies only to closed string propagators) of the propagators. Points in this neighborhood project isomorphically to the uncompactified moduli space ${\cal M}_{b,\{m_i\}}^{g,n}$ and in particular to each connected components ${}_I{\cal M}_{b,\{m_i\}}^{g,n}$. This proves that $[\pi_*{}_I\overline{F}_{b,\{m_i\}}^{g,n}]$ indeed is the fundamental class in $H_{\text{dim}({\cal M}_{b,\{m_i\}}^{g,n})}\left({}_I\overline{\cal M}_{b,\{m_i\}}^{g,n}; \partial{}_I\overline{\cal M}_{b,\{m_i\}}^{g,n}\right)$, and thus the cycle $[\pi_*\overline{F}_{b,\{m_i\}}^{g,n}]$ is the fundamental class in $H_{\text{dim}({\cal M}_{b,\{m_i\}}^{g,n})}\left(\overline{\cal M}_{b,\{m_i\}}^{g,n}; \partial\overline{\cal M}_{b,\{m_i\}}^{g,n}\right)$ we defined above.

Note that the reason why we considered the homology relative to the boundary rather than the homology over the compactified moduli space is due to the infinite length open string propagator and also infinite length closed string propagator connecting to a disk with a bulk puncture. Unlike the other infinite length closed string propagators which do not contribute to the boundary of the chain (because it becomes codimension two as both length and twist parameters disappear), the infinite length open string propagator and also infinite length closed string propagator connected to a disk with a bulk puncture give contributions to the boundary of the chain. This is because the moduli for the open string propagator is one-dimensional and that for closed string propagator connected to a disk with a bulk puncture is also one-dimensional due to the conformal Killing vector. Therefore, the chain under consideration is not a cycle in $H_{\text{dim}({\cal M}_{b,\{m_i\}}^{g,n})}\left(\overline{\cal M}_{b,\{m_i\}}^{g,n}\right)$, even though it is a cycle in the relative homology $H_{\text{dim}({\cal M}_{b,\{m_i\}}^{g,n})}\left(\overline{\cal M}_{b,\{m_i\}}^{g,n}; \partial\overline{\cal M}_{b,\{m_i\}}^{g,n}\right)$. It may seem to suggest that in the actual computation of string amplitudes, total derivatives in the moduli integration give nonzero contributions and thus BRST exact states do not decouple from on-shell amplitudes. However, as long as the background solves the string field equations of motions, Ward identities derived in \cite{Moosavian:2019ydz} imply that null states decouple from on-shell amplitudes, regardless of the presence of boundary of moduli spaces\footnote{We thank Ashoke Sen for discussions on null-state decoupling in open-closed string field theory.}.

We have the following result implying that finding the solution to the geometric master equation (\ref{geometricmaster}) is all we need

\begin{theo}\label{fundamentalhomology}
Feynman diagrams built by solutions to (\ref{geometricmaster}) represent a fundamental class in the homology relative to the boundary when pushed forward to the moduli space.
\end{theo}

In \cite{Costello:2019fuh}, the existence and uniqueness of the solutions to (\ref{geometricmaster}) were also discussed for the closed string vertices. We will not attempt to prove the analogous statements for open-closed vertices, but we believe that it should be a straightforward generalization, with a caveat that the boundary of the Deligne-Mumford compactified moduli space may give rise to subtleties.

\section{Bordered hyperbolic hexagon surfaces}\label{hyperbolicsurfaces}
In this section, we collect mathematical ingredients relevant to the construction of open-closed hyperbolic string vertices. In the case of closed hyperbolic string vertices, the relevant objects were hyperbolic bordered Riemann surfaces \cite{Costello:2019fuh}, where the borders correspond to closed string punctures. Now, we also allow for boundary borders and boundary geodesic sides which will correspond to open string punctures. Therefore our objects of interest are bordered hyperbolic surfaces. Note that these are more general than what is usually called bordered hyperbolic Riemann surfaces i.e. the latter is a proper subset of the former. For instance, the boundary components for bordered hyperbolic surfaces are piecewise geodesic, while those for bordered hyperbolic Riemann surfaces are smooth closed geodesics.

It will turn out that bordered hyperbolic surfaces are too general for our purposes. Thus, we will restrict to bordered hyperbolic hexagon surfaces (BHHS), which is a subset of what is sometimes called hyperbolic surfaces with boundaries and right angles in relevant literatures. Basic building blocks for BHHS are right-angled hexagons, which we describe in detail now.

\begin{figure}[h!]
\centering
\includegraphics[width=1.0\textwidth]{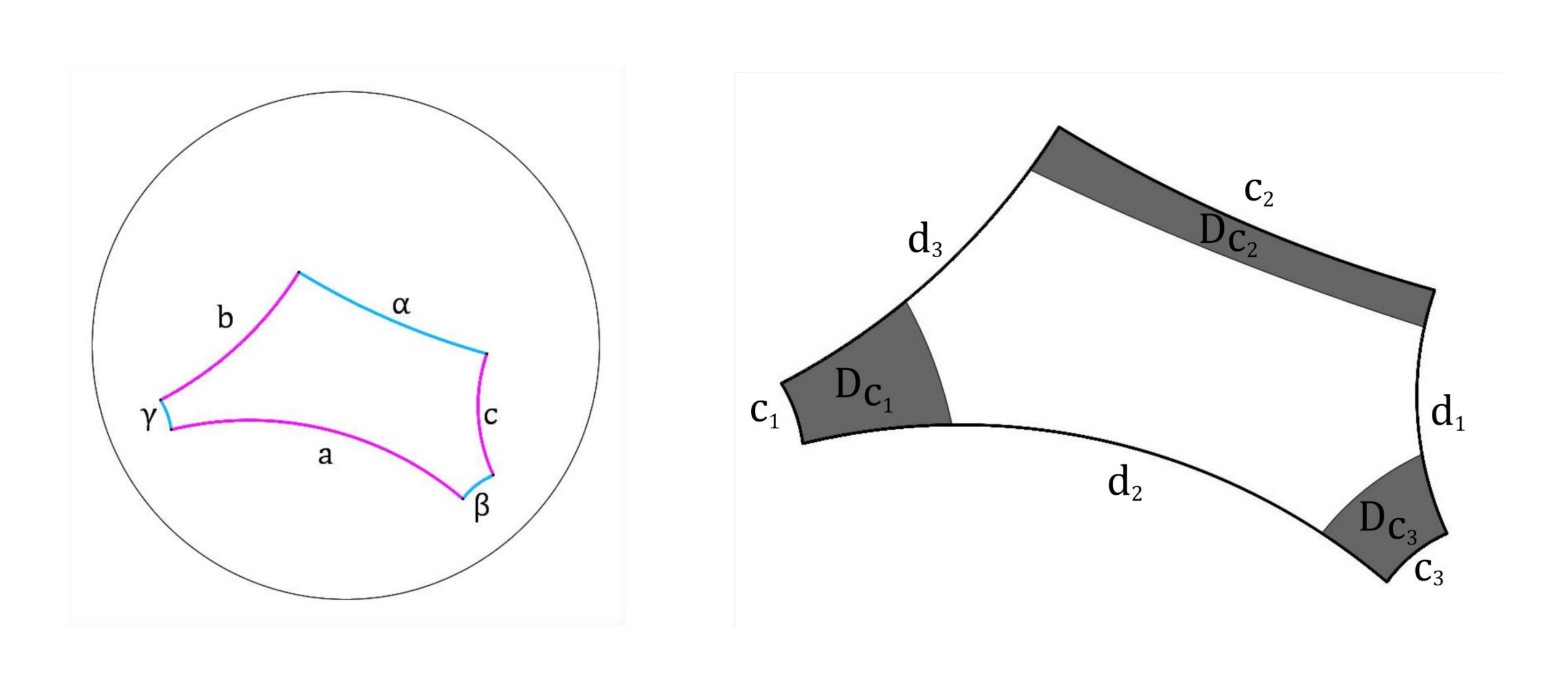}
\caption{Right-angled hexagon drawn on a Poincare disk with three $p$-sides $a, b, c$ (in purple) and three $b$-sides $\alpha, \beta, \gamma$ (in blue), and its half-collars (gray region on the right). Collar theorems are based on the fact that hexagon half-collars do not overlap.}\label{fig:hexagon}
\end{figure}

\subsection{Right-angled hexagons and Y-pieces}
Much of the discussions in this subsection closely follow \cite{Buser1992GeometryAS}. A right-angled hexagon consists of six geodesic sides in the hyperbolic upper-half plane/Poincare disk, with angles between neighboring sides all being right angles, as drawn in Figure \ref{fig:hexagon}. Lengths of three non-consecutive sides determine a right-angled hexagon completely and the following trigonometric identity holds (among several other identities)
\ie\label{trigid}{}
\cosh(c)=\sinh(a)\sinh(b)\cosh(\gamma)-\cosh(a)\cosh(b).
\fe

We introduce half-collars of hexagons, whose properties are basic building blocks for all collar theorems to be discussed later.

\begin{theo}\label{hexagoncollar}{\normalfont (Proposition 3.1.8 in \cite{Buser1992GeometryAS})}
Given a right-angled hexagon $H$ with sides $c_1,d_3,c_2,d_1,c_3,d_2$ labeled in order (say clockwise), define three half-collars associated to $c_i$'s
\ie\label{halfcollar}
D_{c_i}=\{x\in H|\sinh(\text{\normalfont dist}(x,c_i))\sinh{c_i}\leq1 \},
\fe
where {\normalfont dist}$(x,c_i)$ stands for the shortest hyperbolic distance between a point $x$ and a side $c_i$, and $c_i$ appearing as a number (as in $\sinh{c_i}$) stands for its length. Then, three half-collars $D_{c_1,c_2,c_3}$ do not overlap with each other. In particular, the collar $D_{c_i}$ does not overlap with the opposing side $d_i$.
\end{theo}

This is described in Figure \ref{fig:hexagon}. Of course, there is nothing special about choosing half-collars around $c_i$'s, and we could have considered half-collars around $d_i$'s instead and they will not overlap with each other.

In order to build more general surfaces of interest, we introduce labelings of the sides of hexagons.

\begin{define}
A labeled right-angled hexagon is a right-angled hyperbolic hexagon with three non-consecutive sides labeled as $p$-sides and the other three non-consecutive sides labeled as $b$-sides.
\end{define}

We will later attach semi-infinite strips to $p$-sides to turn them into punctures on the boundary component corresponding to open string insertions, while $b$-sides will remain as part of the boundary component. From here on, by a hexagon, we mean a right-angled hexagon with $p$ and $b$-side labelings, unless specified otherwise.

\begin{figure}[h!]
\centering
\includegraphics[width=1.0\textwidth]{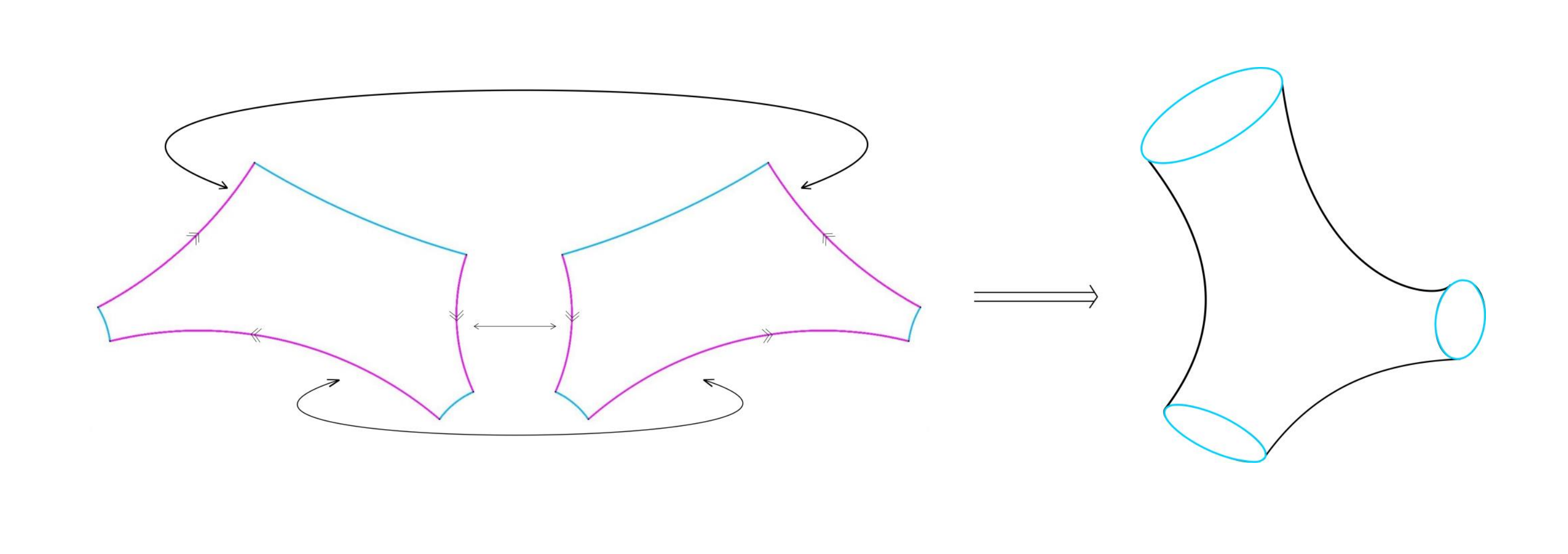}
\caption{A Y-piece obtained by gluing three $p$-sides of two identical hexagons. The resulting Y-piece has three borders and no $p$-sides.}\label{fig:ypiece}
\end{figure}

Gluing two identical hexagons along three $p$-sides generates a Y-piece as illustrated in Figure \ref{fig:ypiece}. It corresponds to a three-bordered sphere, where all borders consist completely of $b$-sides. By construction, a Y-piece is completely determined by lengths of three borders. The conventional hyperbolic bordered Riemann surfaces all can be formed by plumbing Y-pieces together, where plumbing is done by identifying two borders of same lengths taking into account the twist angle.

\subsection{$p$-side gluing of hexagons and BHHS}
Now, we construct more general surfaces by gluing $p$-sides of hexagons, with the requirement that orientations of $b$-sides are preserved. We already saw an example of $p$-side gluing of two hexagons, which is a Y-piece. As another example, for a hexagon which has two $p$-sides of the same lengths, gluing of these two sides will result in an annulus with one $p$-side.

\begin{figure}[h!]
\centering
\subfloat{
\includegraphics[width=0.40\textwidth]{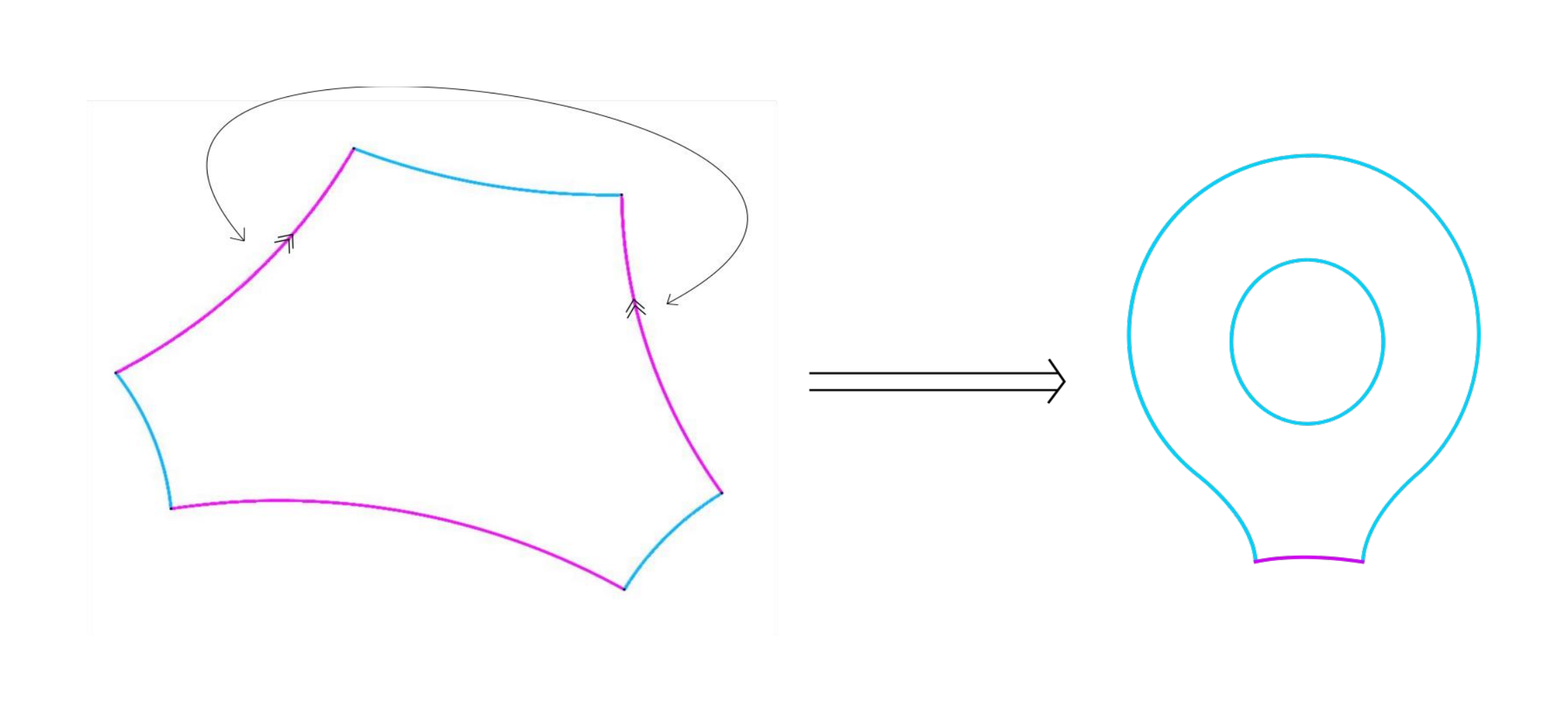}
}
~~~~~~~\subfloat{
\includegraphics[width=0.55\textwidth]{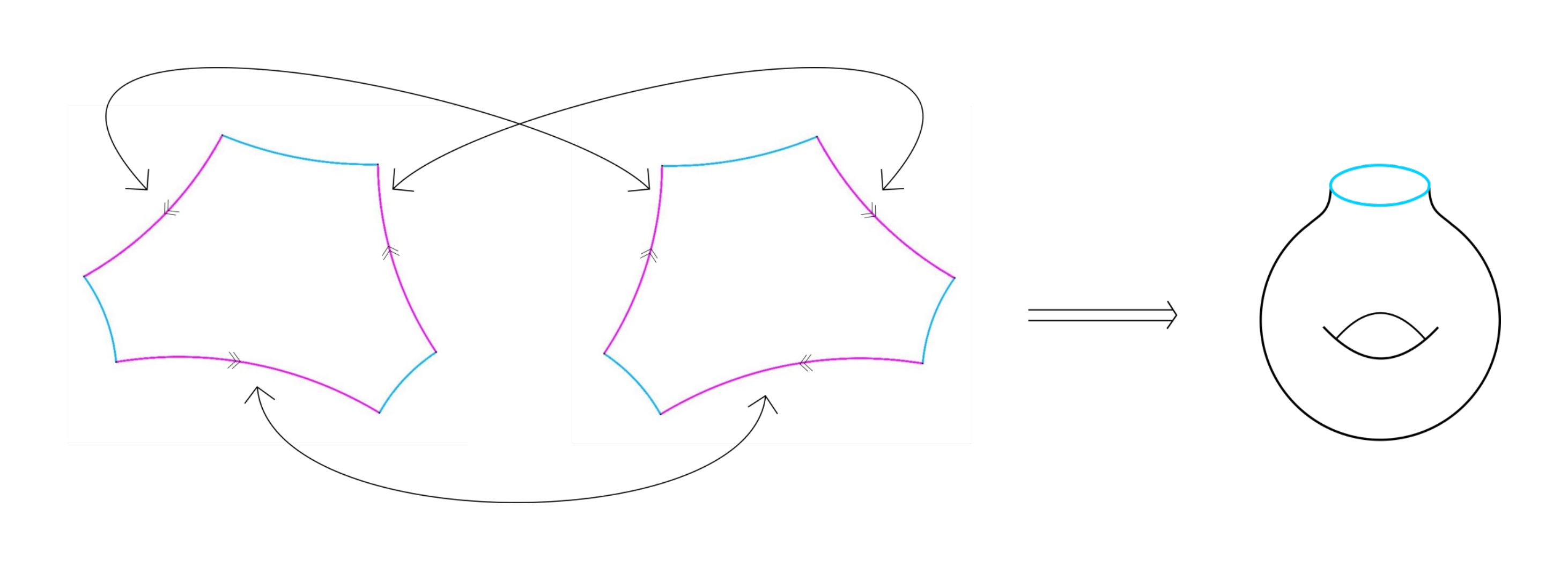}
}
\\
\subfloat{
\includegraphics[width=1\textwidth]{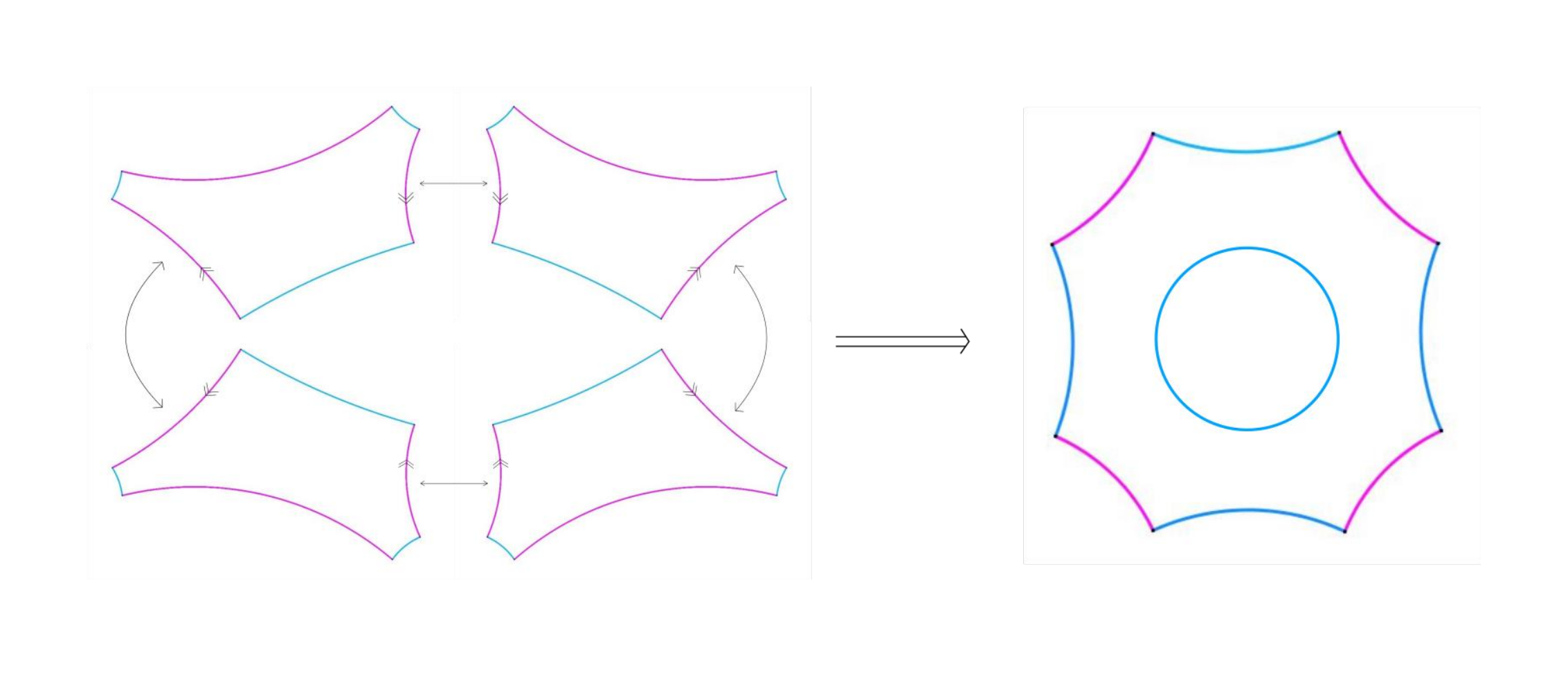}
}
\caption{Examples of BHHS constructed by $p$-side gluing. Top-left is an annulus with one $p$-side, top-right is an one-bordered torus, and the bottom is an annulus with four $p$-sides on one boundary component. $p$-sides are colored in purple while $b$-sides and borders are colored in blue.}\label{fig:BHHS}
\end{figure}

In general, we consider surfaces formed by gluing pairs of $p$-sides of same lengths preserving the orientation of $b$-sides, given some number of hexagons. The resulting hyperbolic surface defines a BHHS. 

\begin{define}
Bordered hyperbolic hexagon surfaces (BHHS) are connected surfaces obtained by gluing pairs of $p$-sides of same lengths of hexagons, where the gluing preserves the orientation of the $b$-sides of hexagons.
\end{define}

We will declare that glued $p$-sides are no longer $p$-sides of the BHHS and only the unglued $p$-sides will remain as its $p$-sides. For example, a Y-piece does not have any $p$-sides as all of the $p$-sides of two hexagons are glued with each other. Several examples are depicted in Figure \ref{fig:BHHS}. The definition for BHHS given here is a constructive one. One can of course start by specifying the notion of moduli space and work backwards to see how hexagons arise. We will indeed discuss such a notion in section \ref{graftingsection} after we further restrict to a more special subset of BHHS.

When a simple smooth closed boundary geodesic is formed under $p$-side gluing (which is possible due to right-angledness of hexagons), it consists only of $b$-sides, and we will call it a border. Also, multiple $b$-sides of hexagons may be smoothly connected to form a larger geodesic side which is not a border. In such cases, we will call the maximally smoothly connected $b$-sides as a single $b$-side of the BHHS, which then should neighbor $p$-sides. Therefore, $b$-sides of a BHHS are geodesic sides of non-border boundary components (in topological sense) which are not $p$-sides. A general BHHS will then carry some number of $p$-sides, some number of $b$-sides, and some number of borders. By construction, any BHHS allows for a hexagon decomposition. Of course, there may be different hexagon decompositions of a given BHHS. This is analogous to different possible pants decompositions of a given bordered Riemann surface. Note that the notion of $p$ and $b$-sides of a BHHS defined above is independent of such different hexagon decompositions. In summary,

\begin{define}
Given a BHHS $S$, consider a hexagon decomposition of $S$. A border of $S$ is a simple smooth closed boundary geodesic of $S$. $p$-sides of $S$ are unglued $p$-sides of hexagons and $b$-sides of $S$ are boundary geodesic sides which are not $p$-sides of $S$ and not borders of $S$.
\end{define}

We are not considering plumbing of borders of same lengths at this point. If one allows for such plumbing in addition, the result still will be a BHHS i.e. it allows for hexagon decompositions, except for the case where the resulting surface has no boundary components, which is nothing but hyperbolic Riemann surfaces without borders with genus greater than or equal to two. Later when we discuss open-closed string vertices, such plumbing of borders will indeed be considered. But for the purpose of discussing a BHHS, such plumbing is not necessary.

\subsection{Collar theorems for BHHS}
Since a BHHS is decomposable into hexagons, the collar theorem for a hexagon (Theorem \ref{hexagoncollar}) has a straightforward generalization to BHHS. We first introduce important open geodesics of interest.

\begin{define}
Given a BHHS $S$, a $p$-geodesic of $S$ is a simple (i.e. no self-intersection) nontrivial (i.e. not homotopic to a point) open geodesic satisfying the following conditions:
\\
i) its endpoints are on either $b$-sides or borders of $S$,
\\
ii) it is the shortest among its homotopy class, where the homotopy allows endpoints to glide along a given $b$-side or border of $S$ (but cannot glide to the other sides).
\end{define}

Given $b$-sides or borders where the endpoints may glide along, these $p$-geodesics are unique among the homotopy class and called perpendiculars since they end on $b$-sides or borders at right angles (Theorem 1.5.3 in \cite{Buser1992GeometryAS} which is also briefly reviewed in Appendix \ref{hyp}). Note that there may be multiple homotopically different $p$-geodesics ending on the same $b$-sides or borders. By definition, all $p$-sides are also $p$-geodesics.

To find a hexagon decomposition of a BHHS, one simply draws a maximal set of pairwise disjoint $p$-geodesics and cut along them, where they become $p$-sides of hexagons after cutting. This works because $p$-geodesics meet $b$-sides or borders at right angles, always preserving the right-angledness and neighborness between $p$ and $b$-sides. In the intermediate steps of cutting, $(2n)$-gons arise with $n\geq3$. No odd-gons appear because they do not respect the notion of $p$ and $b$-sides, and rectangles do not appear because there is no right-angled rectangle in hyperbolic geometry. All $(2n)$-gons with $n\geq3$ can be decomposed into hexagons by cutting along $p$-geodesics.

Given a BHHS $S$ and a $p$-geodesic $\gamma$ on $S$, the collar $P_{\gamma}$ corresponding to it is defined as
\ie
P_{\gamma}=\{x\in S|\sinh(\text{dist}(x,\gamma))\sinh{\gamma}\leq1 \},
\fe
where again, $\gamma$ appearing as a number means its length. Note that when $\gamma$ is a $p$-side, one gets a half-collar in the sense that the width of the collar is a half of that of the collar of a $p$-geodesic of the same length which is not a $p$-side. We have our first collar theorem for BHHS.

\begin{theo}\label{collar1}
Consider a BHHS $S$ and pairwise disjoint $p$-geodesics $\gamma_i$ on $S$. Their collars $P_{\gamma_i}$ do not overlap. Also, the only borders or $b$-sides having nonzero overlap with $P_{\gamma_i}$ are the ones on which $\gamma_i$ ends. Each $P_{\gamma_i}$ is homeomorphic to a strip.
\end{theo}
The basic idea of the theorem is that there is a hexagon decomposition of $S$ where the $p$-geodesics will become $p$-sides of hexagons. Then, the collar theorem of a hexagon, Theorem \ref{hexagoncollar}, implies the above collar theorem for a BHHS. This is along the line of analogous collar theorems for closed geodesics for bordered Riemann surfaces (Theorem 4.1.1 in \cite{Buser1992GeometryAS}).

The second collar theorem considers collars of $b$-sides and borders. Given $\gamma$ which is either a $b$-side or a border on a BHHS $S$, we define the half-collars (since $b$-sides and borders belong to boundary components) as
\ie\label{bcollar}
B_{\gamma}=\{x\in S|\sinh(\text{dist}(x,\gamma))\sinh{\gamma}\leq1 \}.
\fe
When $\gamma$ is a border, the above definition has extra factor of 2 compared to collars in the literature of bordered hyperbolic Riemann surfaces, where $\sinh(\text{dist}(x,\gamma))\sinh(\gamma/2)\leq1$ is used instead. Thus, the half-collars of borders on BHHS are thinner than those of hyperbolic bordered Riemann surfaces. This is because in the latter case, surfaces are built using Y-pieces, which is gluing of two identical hexagons, while for the case of BHHS, a border in a hexagon decomposition may consist of any number of hexagon $b$-sides, including one.

There is a further important point about this. Let $\gamma$ be a $b$-side or a border of a BHHS. In a hexagon decomposition, $\gamma$ generically consists of multiple $b$-sides of hexagons, say $b_1,b_2,...,b_n$ for some $n\geq1$. The length of $\gamma$ is the sum of lengths of $b_i$'s, which implies that the half-collar width for $\gamma$ and half-collar widths for $b_i$'s are different unless $n=1$; they are different even among $b_i$'s of different lengths. However, since the width of a half-collar increases as the length of the associated side decreases, the half-collar of $\gamma$ defined as above is included in the union of the hexagon half-collars of $b_i$'s. Therefore, we have our second collar theorem.

\begin{theo}\label{collar2}
Given a BHHS $S$, consider all $b$-sides $\gamma_i$ and borders $\mu_I$ on $S$. Their half-collars $B_{\gamma_i}$ and $B_{\mu_I}$ do not overlap. Also, each $B_{\gamma_i}$ is homeomorphic to a strip and each $B_{\mu_I}$ is homeomorphic to an annulus.
\end{theo}

One significant point of collar theorems is that any curve passing through a collar has to have a length greater than the width of the collar. This will be used extensively to construct open-closed hyperbolic string vertices in later sections.

\subsection{Grafting: from BHHS to bordered Riemann surfaces with punctures}\label{graftingsection}
In this subsection, we describe how to go from BHHS to surfaces appearing in open-closed string field theory. In order to do so, we introduce one more label to our BHHS. Given a BHHS, say there are $n$ number of borders. We pick a subset of the borders and label them as $c$-borders ($c$ standing for closed strings), and label all the other borders as $b$-borders ($b$ standing for boundaries). We call such labeled BHHS as $c$-labeled BHHS.

\begin{define}
A $c$-labeled BHHS is a BHHS where all borders are labeled either $b$ or $c$. We call them $b$-borders and $c$-borders.
\end{define}

Thus, for a given BHHS with $n$ borders, we generate at maximum $2^n$ $c$-labeled BHHS. For example, for a Y-piece with two of borders carrying the same lengths which are different from that of the third border, giving a single $c$-label to either of two same-length borders will result in the same $c$-labeled BHHS.

For a $c$-labeled BHHS, we will redefine $p$-geodesics. Among the $p$-geodesics of unlabeled BHHS, after $c$-labeling, we will throw away the ones whose either of endpoints lie on the $c$-border.

\begin{define}
Given a $c$-labeled BHHS $S$, a $p$-geodesic of $S$ is a simple nontrivial open geodesic satisfying the following conditions:
\\
i) its endpoints are on either $b$-sides or $b$-borders of $S$,
\\
ii) it is the shortest among its homotopy class.
\end{define}
Again, such a geodesic is unique in its homotopy class by the condition that it is the shortest, and is also the unique common perpendicular to $b$-sides or $b$-borders on which it ends in its homotopy class. We further restrict to the following $c$-labeled BHHS with markings.
\begin{define}
Given a pair of positive real numbers $L_o$ and $L_c$, the associated open-closed hyperbolic surfaces is a set defined as:
\\
$H_{L_o,L_c}$ = \{$c$-labeled BHHS whose $p$-sides all have lengths $L_o$, $c$-borders all have lengths $L_c$,
$~~~~~~~~~~~~$ and $p$-sides and $c$-borders are marked\}.
\end{define}

So far we have not defined the notion of equivalance of elements in $H_{L_o,L_c}$. This is the content of the moduli space. The notion of moduli space associated with $H_{L_o,L_c}$ is specified by its genus $g\geq0$, number $n\geq0$ of $c$-borders, number $b\geq0$ of non-$c$-border boundary components (in topological sense), number $m_i\geq0$ of $p$-sides on the $i$-th non-$c$-border boundary component, and markings for $p$-sides and $c$-borders. We call this moduli space ${\cal{M}}_{b,\{m_i\}}^{g,n}(L_o,L_c)$. Note that our construction of BHHS implies that the possible values of $g, n, b$ and $m_i$ are the same as those of open-closed string vertices (\ref{moduliconditions}), with the exception of a disk with a bulk puncture, an annulus with no punctures, and the cases $n=b=0$ with $g\geq2$ which are closed Riemann surfaces.

Given the numbers $g,n,b,\{m_i\}$ in (\ref{moduliconditions}), excluding two cases $\{g=0,n=1,b=1,m=0\}$ and $\{g=0,n=0,b=2,\{m_1=0,m_2=0\}\}$ which we will separately deal with later, the precise formulation of the moduli space goes as follows. We consider an orientable connected surface $S$ of genus $g$ and $n+b$ number of boundaries. Teichm\"{u}ller space ${\cal T}_{b,\{m_i\}}^{g,n}(L_o,L_c)(S)$ is given by the set of all hyperbolic metrics on the surface satisfying: i) $n$ boundaries (which we call $c$-borders) are simple smooth closed geodesics of lengths $L_c$ and marked, ii) other $b$ number of boundaries, say $b_i$ for $i=1...,n$, each consists of $2m_i$ number of simple geodesic sides which meet neighboring sides at right angles for $m_i>0$, while $b_i$ with $m_i=0$ is a simple smooth closed geodesic which we call a $b$-border, and iii) for each of $b_i$ with $m_i\neq0$, one nonconsecutive half of the sides are called $p$-sides all of lengths $L_o$ which are marked, and the other nonconsecutive half of them are called $b$-sides. As in the case of $c$-labeled BHHS, we define a $p$-geodesic to be the unique shortest simple nontrivial open geodesic in its homotopy class with the endpoints on either $b$-borders or $b$-sides of the surface. The mapping class group $\Gamma(S)$ is the quotient of all orientation preserving diffeomorphisms by all diffeomorphisms connected to the identity, where diffeomorphisms must preserve $b,c$-borders, $b,p$-sides, and markings. Then, the moduli space is defined by ${\cal{M}}_{b,\{m_i\}}^{g,n}(L_o,L_c)\equiv{\cal T}_{b,\{m_i\}}^{g,n}(L_o,L_c)(S)/\Gamma(S)$. Surfaces in ${\cal{M}}_{b,\{m_i\}}^{g,n}(L_o,L_c)$ are elements of $H_{L_o,L_c}$ except for the cases $n=b=0$ with $g\geq2$ as already noted above.

Even though closed hyperbolic Riemann surfaces with $g\geq2$ do not belong to $H_{L_o,L_c}$, once we cut along any nontrivial simple closed geodesic and label it $b$-borders, they become elements of $H_{L_o,L_c}$. For these surfaces, the lengths and twist angles of nontrivial simple closed geodesics associated with a pants decomposition provide a natural parameterization of the Teichm\"{u}ller space, which are Fenchel-Nielsen coordinates. Or, one can also cut along a chosen simple closed geodesic and consider the hexagon decomposition of the resulting surface, where the length and twist angle of the closed geodesic and lengths of hexagon $p$-sides provide another coordinates of the Teichm\"{u}ller space. For all the other cases, a hexagon decomposition of an open-closed hyperbolic surface provides the lengths of hexagon $p$-sides as coordinates for the Teichm\"{u}ller space, once the conditions on the lengths of $p$-sides and $c$-borders of the open-closed hyperbolic surface are imposed.

\begin{figure}[h!]
\centering
\includegraphics[width=1.0\textwidth]{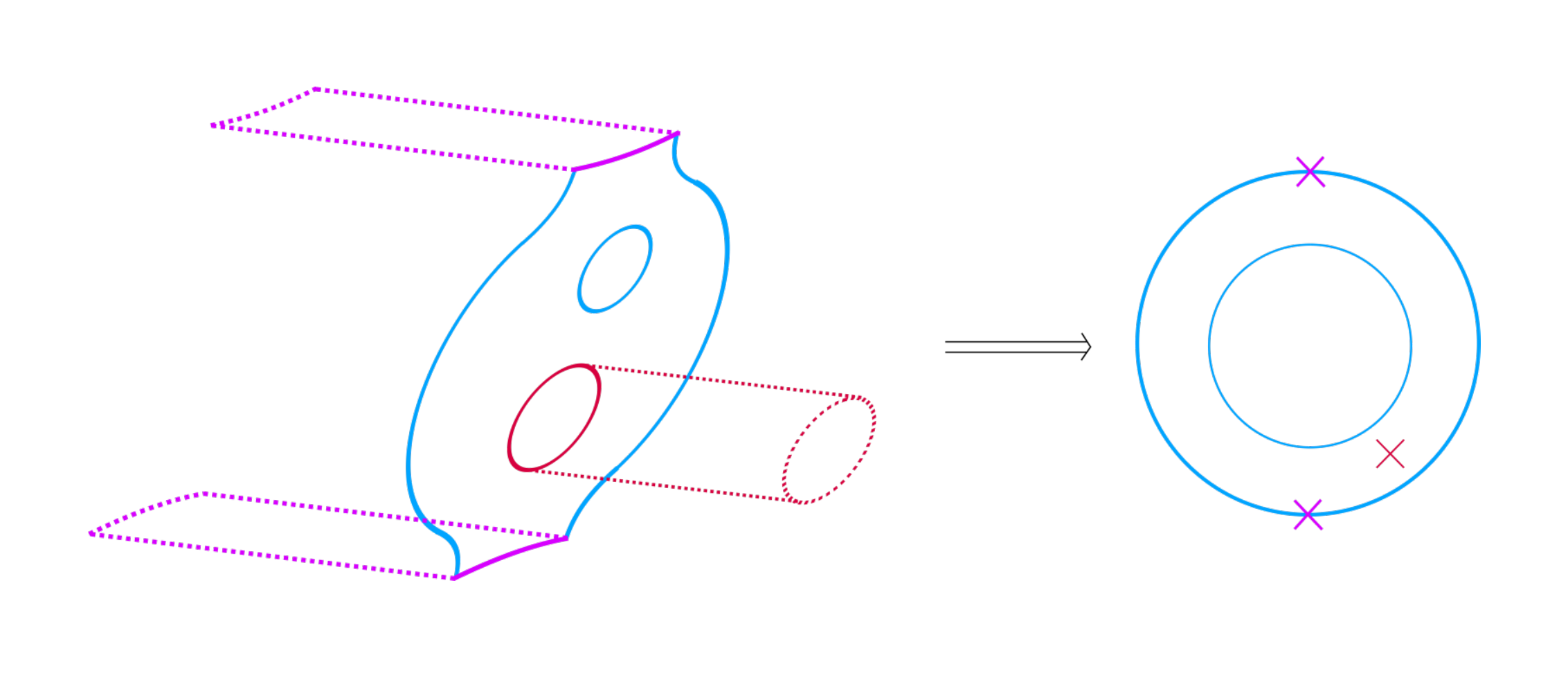}
\caption{Grafting a surface in ${\cal{M}}_{2,\{2,0\}}^{0,1}(L_o,L_c)$. Flat semi-infinite strips of width $L_o$ (dotted purple) are attached to $p$-sides of lengths $L_o$ (solid purple) and a flat semi-infinite cylinder with circumference length $L_c$ (dotted cherry) is attached to a $c$-border of circumference length $L_c$ (solid cherry). The result is an element in ${\hat{\cal P}}_{2,\{2,0\}}^{0,1}$, which upon projection gives an element in ${\cal{M}}_{2,\{2,0\}}^{0,1}$ shown on the right. Boundaries are colored in blue.}\label{fig:grafting}
\end{figure}

Now, we introduce a map from ${\cal{M}}_{b,\{m_i\}}^{g,n}(L_o,L_c)$ to bordered Riemann surfaces with marked bulk (closed string) and boundary (open string) punctures, which are basic objects of open-closed string field theory. The map is called "grafting," and as the name suggests, we will graft other surfaces to those in ${\cal{M}}_{b,\{m_i\}}^{g,n}(L_o,L_c)$. Given an element of ${\cal{M}}_{b,\{m_i\}}^{g,n}(L_o,L_c)$, to every $p$-side, we glue a flat semi-infinite strip of width $L_o$, and to every $c$-border, we glue a flat semi-infinite cylinder with circumference length $L_c$, as described in Figure \ref{fig:grafting}. A grafted semi-infinite strip introduces a puncture on the boundary (open string puncture), while a grafted semi-infinite cylinder introduces a puncture in the bulk (closed string puncture). Note that gluing is done isometrically, so the metric is continuous across the glued parts, even though it is not smooth; for example, the curvature jumps.

Since grafting results in bordered Riemann surfaces with marked bulk and boundary punctures together with specific metric on it, it is a map to the total space over the moduli space of such punctured Riemann surfaces. Thus, grafting is a map (following the notations in \cite{Costello:2019fuh})
\ie
\text{gr}'_{\infty}~:~~{\cal{M}}_{b,\{m_i\}}^{g,n}(L_o,L_c)~\rightarrow~{\hat{\cal P}}_{b,\{m_i\}}^{g,n}.
\fe
Composing with the projection map $\pi:{\hat{\cal P}}_{b,\{m_i\}}^{g,n}\rightarrow{\cal{M}}_{b,\{m_i\}}^{g,n}$, we also get the following map.
\ie\label{graft}
\text{gr}_\infty\equiv\pi\circ\text{gr}'_\infty~:~~{\cal{M}}_{b,\{m_i\}}^{g,n}(L_o,L_c)~\rightarrow{\cal{M}}_{b,\{m_i\}}^{g,n}.
\fe
Note that in cases $n=0$ and $m_i=0$ for all $i$, where there is nothing to graft, the hyperbolic surfaces already provide chains in the total space defined by the hyperbolic structures. 

In the case of hyperbolic bordered Riemann surfaces where $b=0$ in the above, such grafting map gr${}_\infty$ was shown to be a homeomorphism \cite{mondello2008riemann,scannell1998grafting}. We will not try to prove that (\ref{graft}) is a homeomorphism for general $b$ and $\{m_i\}$, but we believe it should be true and is a straightforward generalization of the story for hyperbolic bordered Riemann surfaces. So we will assume that the grafting map indeed is a homeomorphism.

\section{Open-closed hyperbolic string vertices}\label{hyperbolicvertices}
In this section, we construct hyperbolic open-closed string vertices using $H_{L_o,L_c}$ and grafting. We will show that with appropriate conditions on $L_o$ and $L_c$, we get the solution to the open-closed geometric master equation (\ref{geometricmaster}).

\subsection{Critical length and open-closed vertex region}
Essentially, the region of $L_o$ and $L_c$ for vertices will be such that collars are wide enough to make curves passing through them lengthy enough. Here, we will define such a region without any explanation or motivation. But once we discuss the proof of the open-closed string vertex solutions, it will become clear why we defined such a region.

First, let us define the critical length $L_*\in\mathbb R_+$. It is defined to be the solution to the following equation
\ie\label{critical}
\sinh\left({L_*\over2}\right)\sinh L_*=1.
\fe

\begin{figure}[h!]
\centering
\subfloat{
\includegraphics[width=.7\textwidth]{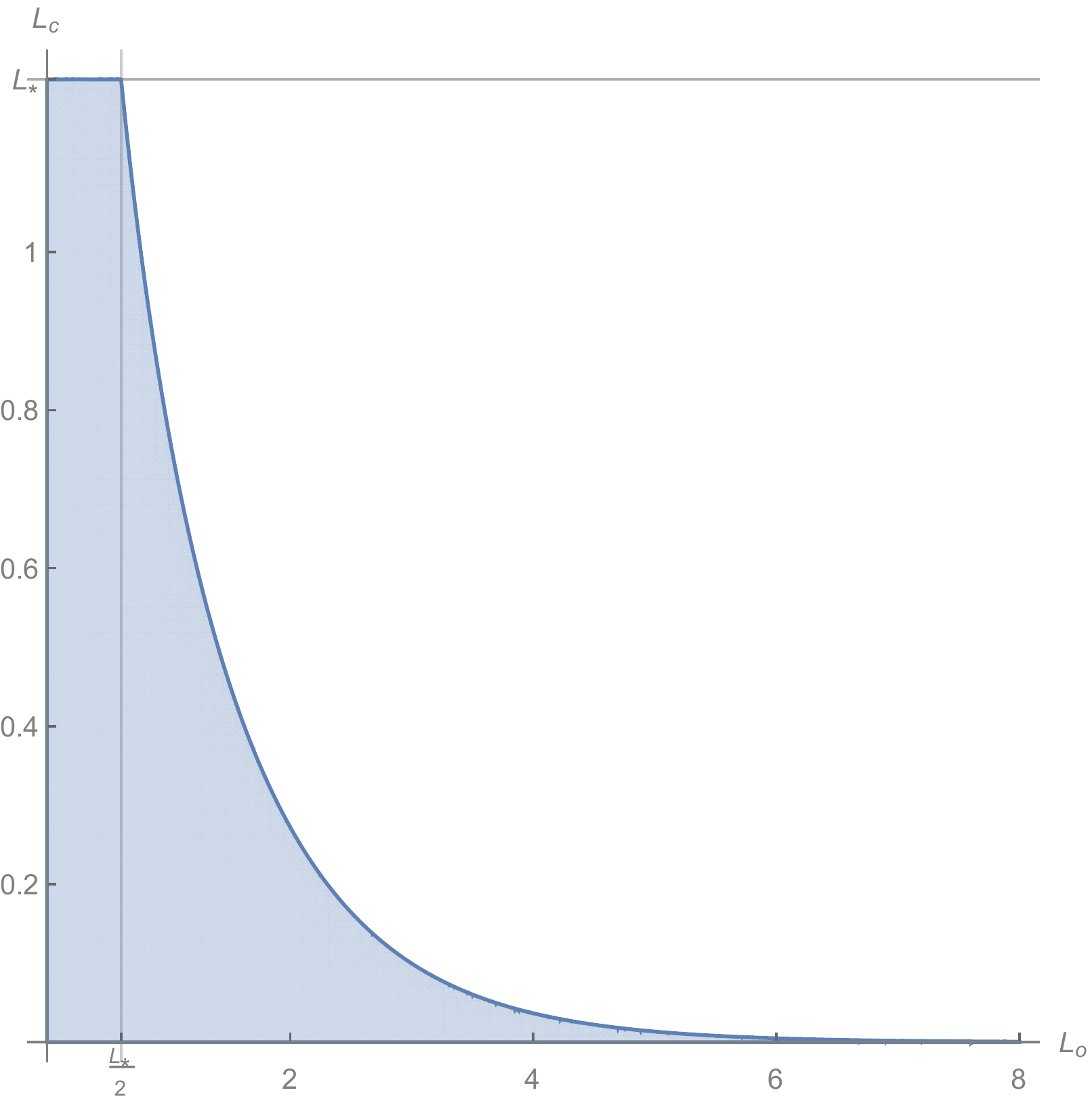}
}
\caption{Region ${\cal R}$ for $L_o$ and $L_c$. For $L_o\geq {L_*\over2}$, the region is given by $\sinh L_c {\sinh L_o}\leq1$.}
\label{fig:regionR}
\end{figure}

Numerical value for $L_*$ is approximately 1.21876.... Note that this critical length is smaller than the critical length defined in \cite{Costello:2019fuh}, where it was $2\sinh^{-1}1=1.76275...$. This is due to different collar widths around nontrivial simple closed geodesics for the case of BHHS and bordered hyperbolic Riemann surfaces, as already discussed below (\ref{bcollar}). We define the following open-closed vertex region of $L_o$ and $L_c$
\ie\label{regionR}
{\cal R}=\left\{(L_o,L_c)\in{\mathbb R}_+^2~\bigg|~0<L_c\leq L_*\text{  and  }\sinh L_c \sinh L_o \leq1 \right\}.
\fe
It is depicted in Figure \ref{fig:regionR}. For $L_o\leq {L_*\over2}$, it is a rectangular region given by $0<L_o\leq {L_*\over2}$ and $0<L_c\leq L_*$. For $L_o\geq {L_*\over2}$, we have a more strict upper bound on $L_c$ given by $\sinh L_c \sinh L_o \leq1$.

Even though it is a trivial result, we record the following which will be used later in the proof for string vertices.
\ie\label{regionR2}
\text{All points in region $\cal R$ satisfy }\sinh L_c\sinh{L_o\over2}\leq1, \text{ and also } \sinh L_o\sinh {L_c\over2}\leq1.
\fe

\subsection{A family of open-closed string vertices}
We first define the subset ${\tilde{\cal V}}_{b,\{m_i\}}^{g,n}(L_o,L_c)$ of the moduli space ${\cal{M}}_{b,\{m_i\}}^{g,n}(L_o,L_c)$, excluding two cases $\{g=0,n=1,b=1,m=0\}$ and $\{g=0,n=0,b=2,\{m_1=0,m_2=0\}\}$, as follows
\ie
{\tilde{\cal V}}_{b,\{m_i\}}^{g,n}(L_o,L_c)\equiv\left\{\Sigma\in{\cal{M}}_{b,\{m_i\}}^{g,n}(L_o,L_c)~\bigg|~\text{sys}(\Sigma)\geq L_c\text{ and }\text{psys}(\Sigma)\geq L_o \right\}.
\fe
Here, sys$(\Sigma)$ is the systole, which is the length of the shortest nontrivial simple closed geodesic in $\Sigma$ which is not a $c$-border. In particular, $b$-borders are included in the consideration of the systole. Also, psys$(\Sigma)$ is the length of the shortest $p$-geodesic of $\Sigma$ which is not a $p$-side. Therefore, surfaces in ${\tilde{\cal V}}_{b,\{m_i\}}^{g,n}(L_o,L_c)$ have no nontrivial simple closed geodesic of length smaller than $L_c$ and no $p$-geodesic of length smaller than $L_o$.

In cases $b=0$ where there is no $b$-border or $b$-side, there is no $p$-geodesic to consider and thus the condition $\text{psys}(\Sigma)\geq L_o$ is empty. These correspond to pure closed string processes and the systolic condition is imposed only on nontrivial simple closed geodesics, as in \cite{Costello:2019fuh}. For $b\geq1$, the surfaces are elements of $H_{L_o,L_c}$ with nontrivial $p$-geodesics to consider generically.

We now define the grafting of these surfaces
\ie
{\cal V}_{b,\{m_i\}}^{g,n}(L_o,L_c)\equiv\text{gr}'_\infty\left({\tilde{\cal V}}_{b,\{m_i\}}^{g,n}(L_o,L_c)\right).
\fe
We also define the two cases which have not been considered so far. The first one is a disk with a bulk puncture
\ie\label{disk1pt}
\tilde{\cal V}_{b=1,\{m=0\}}^{g=0,n=1}(L_o,L_c)\equiv\{\text{a flat circle of circumference length } L_c \text{ with a marking}\}.
\fe
We will declare that the circle is a $c$-border, even though it is not bounding any surface. Then, grafting acting on the circle will simply result in a flat semi-infinite cylinder of circumference length $L_c$. The second one is an annulus without punctures
\ie
{\cal V}_{b=2,\{m_1=0,m_2=0\}}^{g=0,n=0}(L_o,L_c)\equiv\varnothing.
\fe
Now, we have all the ingredients to describe hyperbolic open-closed string vertices.

\begin{theo}\label{vertextheo}
The sets ${\cal V}_{b,\{m_i\}}^{g,n}(L_o,L_c)$ with any given pair $(L_o,L_c)\in{\cal R}$ solve the open-closed geometric master equation (\ref{geometricmaster}).
\end{theo}
The special case $\partial{\cal V}_{2,\{0,0\}}^{0,0}=-{1\over2}\{{\cal V}_{1,\{0\}}^{0,1},{\cal V}_{1,\{0\}}^{0,1}\}$ is trivially satisfied. From here on, we will elaborate on the proof of the general cases. The proof is in two parts. First, we prove that boundary $\partial\cal V$ of the candidate vertex set is contained in $-{1\over2}\{{\cal V},{\cal V}\}-\hbar\Delta\cal V$. Then, we prove the opposite direction that both $-{1\over2}\{{\cal V},{\cal V}\}$ and $-\hbar\Delta\cal V$ are contained in $\partial\cal V$.

The first part is easier than the second part. Boundary of $\cal V$ corresponds to surfaces $\Sigma$ where either a nontrivial non-$c$-border simple closed geodesic in $\Sigma$ becomes of length $L_c$, or a non-$p$-side $p$-geodesic becomes of length $L_o$. In the former case, if the closed geodesic were a $b$-border, then it is obtained by $\{{\cal V}_{b=1,\{m=0\}}^{g=0,n=1},{\chi}\}_c$ where ${\chi}$ is the same surface as $\Sigma$ except that the $b$-border of length $L_c$ in $\Sigma$ is a $c$-border of $\chi$. Such $\chi$ belongs to the vertex set because there are no $p$-geodesics shorter than $L_o$ and no nontrivial simple closed geodesics shorter than $L_c$, since $\Sigma$ satisfies the same condition by assumption.

If the nontrivial non-$c$-border simple closed geodesic of length $L_c$ is not a $b$-border, we cut along the closed geodesic to produce two new borders of lengths $L_c$ which we declare to be $c$-borders. If the result were two disjoint surfaces, then it belongs to $\{{\cal V,V}\}_c$ since no nontrivial simple closed geodesics are shorter than $L_c$, no $p$-geodesics are shorter than $L_o$, and the two new $c$-borders are both of lengths $L_c$. If the result of cutting were still a connected surface, say $\xi$, again no nontrivial simple closed geodesics are shorter than $L_c$ and no $p$-geodesics are shorter than $L_o$ and the original surface is obtained by acting $\Delta_c$ on $\xi$.

In the case where a non-$p$-side $p$-geodesic becomes of length $L_o$, the argument is similar and it is contained in $\{{\cal V,V}\}_o$ or $\Delta_o\cal V$, where two new sides of lengths $L_o$ arising from cutting are declared to be $p$-sides. This completes the first part of the proof.

The second part of the proof separates into four cases. In each of the cases where we obtain new surfaces from $\{\cal V,V\}$ or $\Delta{\cal V}$, we have to show that there are no new nontrivial simple closed geodesic of length smaller than $L_c$ and no new $p$-geodesic of length smaller than $L_o$. Then the resulting surfaces will be included in $\partial\cal V$.

\begin{figure}[h!]
\centering
\includegraphics[width=1.0\textwidth]{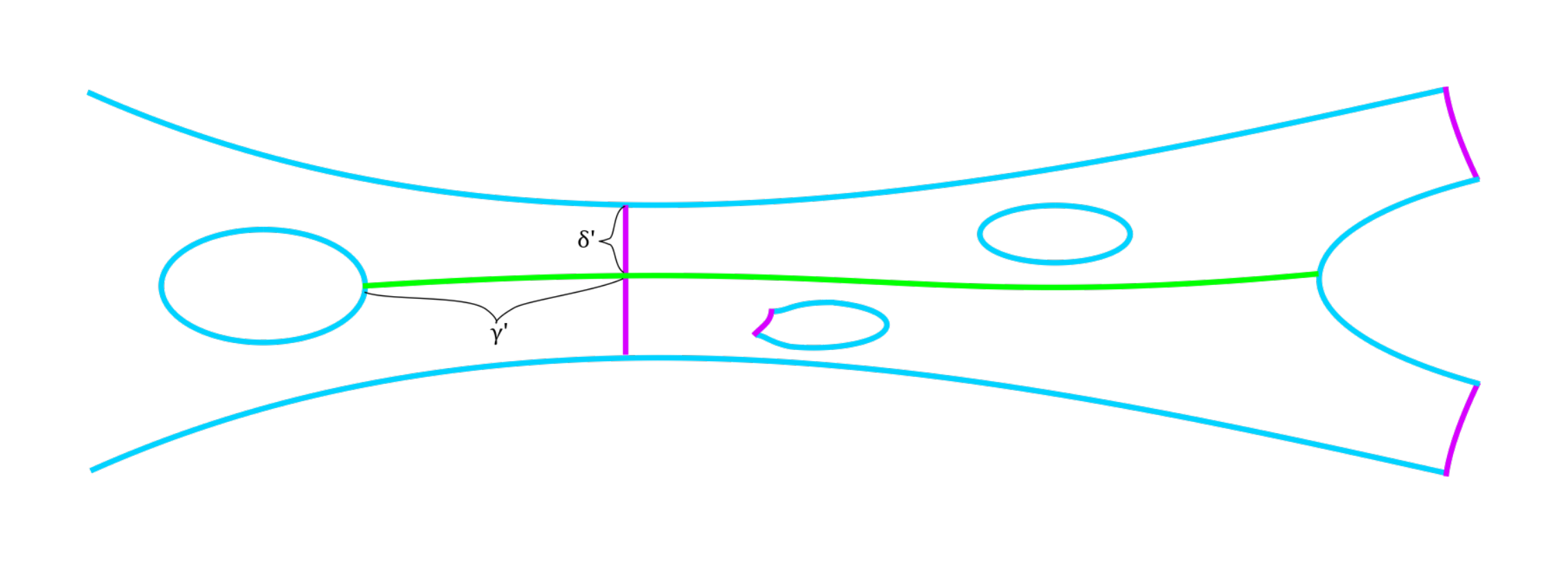}
\caption{Cutting a possible new $p$-geodesic (in green) along the glued $p$-side of length $L_o$ (in purple in the middle). If it were shorter than $L_o$, the union of the curve $\gamma'$ and $\delta'$ has length shorter than $L_o$ leading to a contradiction.}\label{fig:openglue1}
\end{figure}

\begin{case}\label{case1}
$\{{\cal V,V}\}_o$ belongs to $\partial\cal V$.
\end{case}
The new surface is obtained by gluing two $p$-sides of two disjoint surfaces. Any new $p$-geodesics or nontrivial simple closed geodesic for the new surface should pass through the glued $p$-side of length $L_o$. Towards contradiction, assume that there is a new $p$-geodesic $\gamma$ which passes through the glued $p$-side, and it has length smaller than $L_o$. This is depicted in Figure \ref{fig:openglue1}. Note that it may pass through the $p$-side multiple times. We cut $\gamma$ along the glued $p$-side and among possibly many pieces (at least two pieces), there will be two arcs with one endpoint on either a $b$-side or a $b$-border and the other endpoint on the glued $p$-side. At least one of the two arcs should be of length smaller than ${L_o\over2}$. Call it $\gamma'$. The endpoint of $\gamma'$ on the glued $p$-side divides the $p$-side into two segments, so one of them should have length smaller than or equal to ${L_o\over2}$. Call it $\delta'$. We define $\gamma''$ to be the union of two arcs $\delta'$ and $\gamma'$, and then $\gamma''$ has length strictly smaller than $L_o$. But $\gamma''$ is completely contained in one of the two vertices we glued together, and is an arc with both endpoints on $b$-sides or $b$-borders. Thus, in its homotopy class, there will be a $p$-geodesic having length shorter than $L_o$, which is a contradiction. One may wonder about the case where the arc $\gamma'$ is homotopic to a point. In this case, one can always glide the endpoint on $b$-side or $b$-border to the end of the glued $p$-side. Trigonometric identity (\ref{triangleid}) guarantees that such glided arc has shorter length, so such case is not a $p$-geodesic to begin with. Another way to see this is that since a $p$-geodesic is a perpendicular to the $b$-sides or $b$-borders on which it ends, we cannot form a triangle with two sides being $\gamma'$ and $\delta'$, thus again making the curve under consideration not a $p$-geodesic.

\begin{figure}[h!]
\centering
\includegraphics[width=1.0\textwidth]{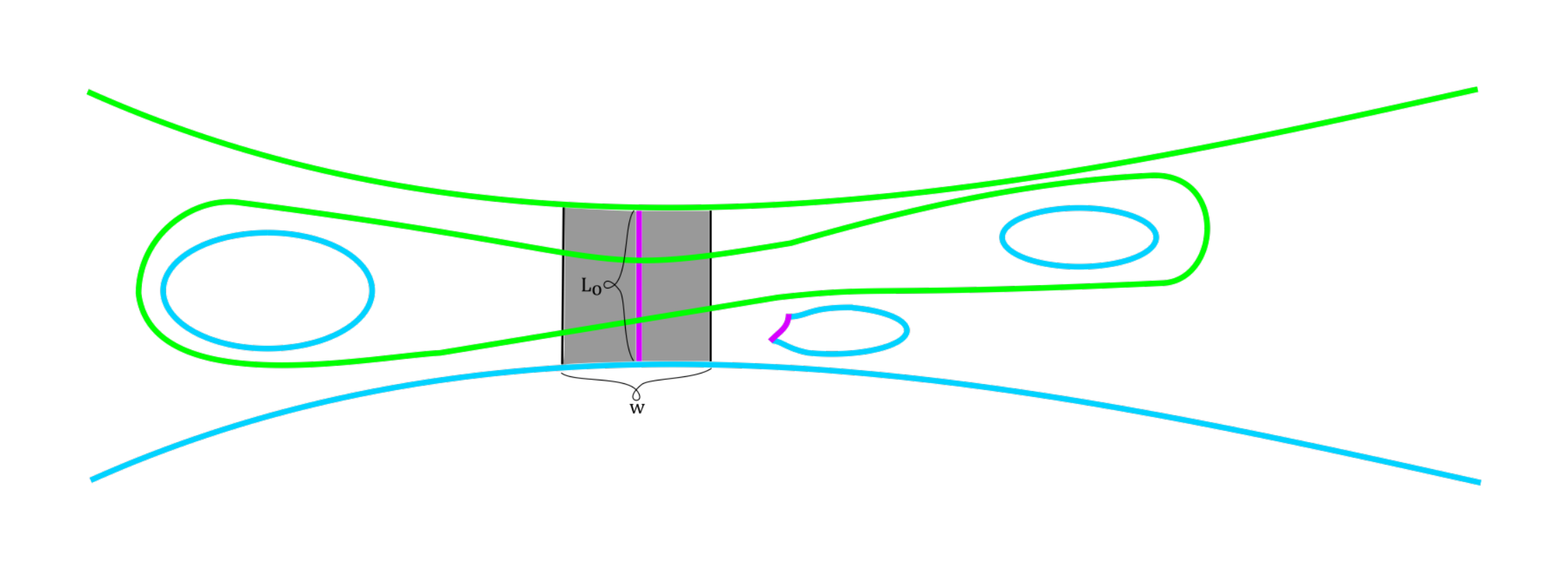}
\caption{Possible new simple closed geodesics (in green) as $p$-sides of lengths $L_o$ (in purple in the middle) are glued. These new closed geodesics must pass through the collar of the $p$-side (in gray).}\label{fig:openglue2}
\end{figure}

\begin{figure}[h!]
\centering
\includegraphics[width=0.8\textwidth]{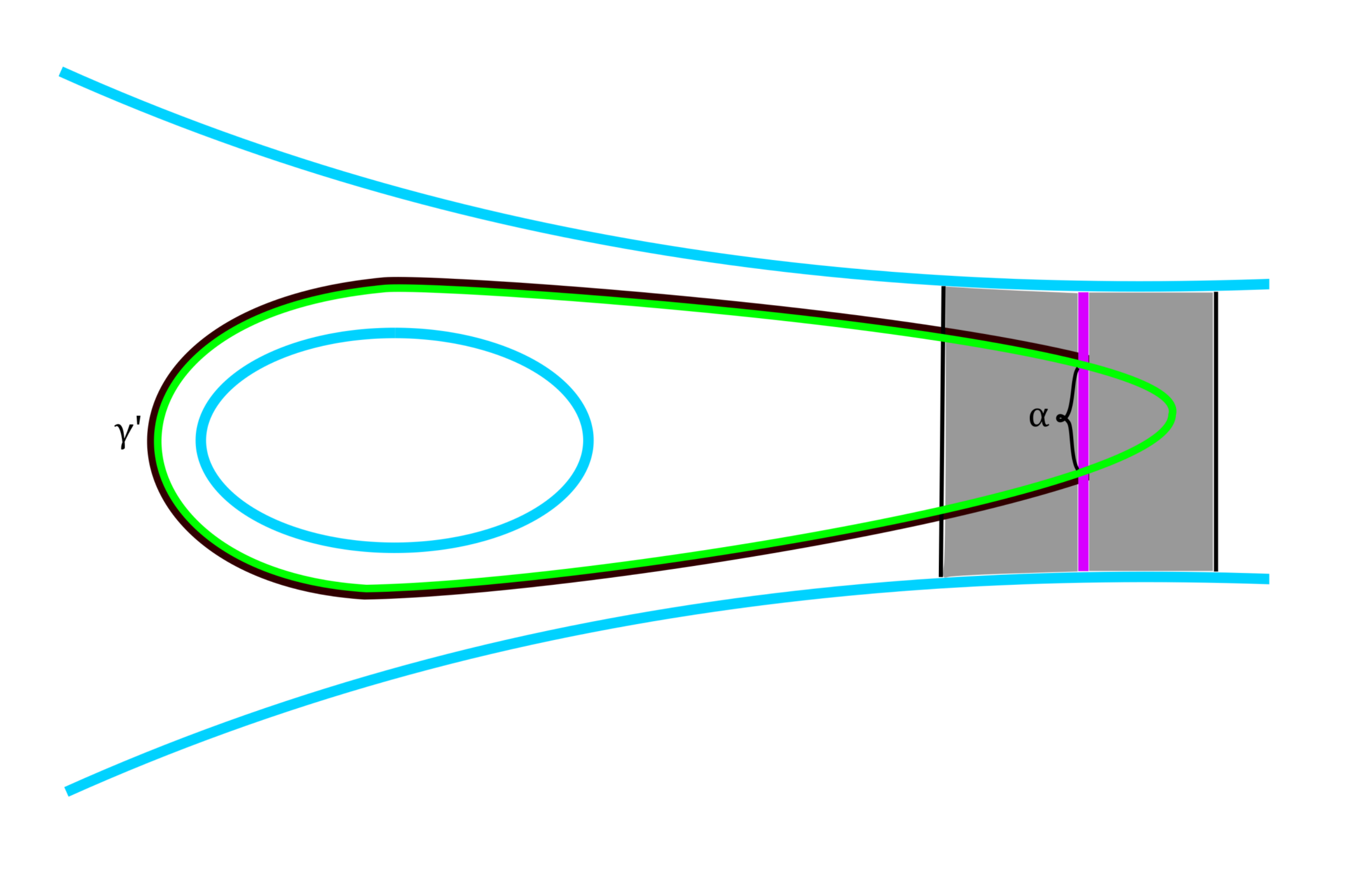}
\caption{A new simple closed geodesic cannot cross only one end of the collar and thus must pass through both ends of the collar. In the above case, the closed curve formed by the union of $\alpha$ and $\gamma'$ is completely contained in the left surface and thus is freely homotopic to a unique closed geodesic which also should be completely contained in the left surface, meaning that the candidate closed geodesic (in green) is not a geodesic.}\label{fig:openglue3}
\end{figure}

Now we discuss new nontrivial simple closed geodesics. Any such closed geodesic should pass through the glued $p$-side. We consider the collar of this $p$-side, which will have width $w$ given by $\sinh{w\over2}={1\over\sinh L_o}$ as in Figure \ref{fig:openglue2}. By Theorem \ref{collar1}, the closed geodesic under consideration cannot be fully contained in the collar and it also has to pass through both ends of the collar. Note that the case like Figure \ref{fig:openglue3} cannot happen. If we cut along the glued $p$-side and look at the closed curve formed by the union of $\alpha$ and $\gamma'$, it is completely contained in one of the two glued surfaces. Theorem \ref{closedgeotehroem} in Appendix \ref{hyp} says that any closed curve is freely homotopic to a unique closed geodesic, meaning that in this case, the unique closed geodesic is also completely contained in one surface. Therefore, the original closed curve under consideration was not a closed geodesic to begin with.

Then, the length $L$ of the closed geodesic has to be greater than $w$: $L>w$. Using the relation between the width of the collar and $L_o$, we get
\ie\label{newclosedopenglue}
\sinh{L\over2}>\sinh{w\over2}={1\over\sinh{L_o}}\geq\sinh{L_c\over2},
\fe
where the last inequality follows from (\ref{regionR2}). Therefore, the closed geodesic has length greater than $L_c$, which is what we wanted to prove.
 
\begin{case}
$\Delta_o{\cal V}$ belongs to $\partial\cal V$.
\end{case}
A new surface is acquired by gluing two $p$-sides of a single surface. We first consider possibly new $p$-geodesics. They have to pass through the glued $p$-side. If their two ends lie on $b$-borders or $b$-sides which are not neighboring either of the two glued $p$-sides, we can assume that such a $p$-geodesic has length shorter than $L_o$ and show that it leads to a contradiction, exactly in the same way as in Case \ref{case1}. If either of the two endpoints lie on a $b$-side neighboring the glued $p$-side, we again cut the curve along the glued $p$-side and consider the piece ending on the neighboring $b$-side. If that piece can be homotopically glided to the glued $p$-side, trigonometric identity for a right-angled triangle (\ref{triangleid}) implies that the arc after the gliding is shorter, thus making the original curve not a geodesic. If not, we can again use the contradiction argument.

The case of a new nontrivial simple closed geodesic can be treated in the exactly same manner as Case \ref{case1}, again illustrated in Figure \ref{fig:openglue2}. Such a closed geodesic should cross both ends of the collar of the glued $p$-side, thus being lengthier than the collar width, which is lengthier than $L_c$ by (\ref{regionR2}) and (\ref{newclosedopenglue}).

\begin{case}\label{case3}
$\{{\cal V,V}\}_c$ belongs to $\partial\cal V$.
\end{case}

We first discuss a seemingly trivial, but very important special case, where one of the vertex is ${\cal V}_{b=1,\{m=0\}}^{g=0,n=1}$. Then, twist-plumbing it to a $c$-border of another surface simply turns the label into a $b$-border, still having length $L_c$. Thus, there are no new nontrivial simple closed geodesics having lengths smaller than $L_c$. However, there are new $p$-geodesics whose at least one of two endpoints lies on the new $b$-border of length $L_c$. Such $p$-geodesics should pass through the half-collar of the $b$-border by Theorem \ref{collar2}. The width of the half-collar $w$ is given by $\sinh w={1\over\sinh L_c}$. By the definition of the vertex region $\cal R$ in (\ref{regionR}), we have $w\geq L_o$, which shows that the new $p$-geodesics are lengthier than $L_o$. This is the only case where a possible new geodesic passes through only the half-collar, rather than a full collar. If this case were absent, the definition of the vertex region ${\cal R}$ in (\ref{regionR}) could have been less restrictive, requiring $L_c\leq L_*$ and $\sinh{L_o}\sinh{L_c\over2}\leq1$ instead.

Now, we discuss all the other general cases where two disjoint surfaces are twist-plumbed along two $c$-borders, say $\gamma$. We first consider new $p$-geodesics. Since they end on $b$-sides or $b$-borders, they should pass through the collar of $\gamma$ by Theorem \ref{collar2}. Since $\gamma$ is of length $L_c$, its collar has width $w$ given by $\sinh{w\over2}={1\over\sinh L_c}$. Then, we have
\ie
\sinh{w\over2}={1\over\sinh{L_c}}\geq\sinh{L_o\over2},
\fe
where the last inequality follows from (\ref{regionR2}). Therefore, we conclude that $w\geq L_o$ and thus new $p$-geodesics all have lengths greater than $L_o$.

For new nontrivial simple closed geodesics, Theorem \ref{collar2} again implies that they should be lengthier than the width of the collar $w$, given by $\sinh {w\over2}={1\over\sinh{L_c}}$. Since (\ref{regionR}) requires $L_c\leq L_*$, the definition of $L_*$ given in (\ref{critical}) implies
\ie
\sinh{L_c}\sinh{L_c\over2}\leq1~~\Rightarrow~~\sinh{L_c\over2}\leq{1\over\sinh{L_c}}=\sinh{w\over2}.
\fe
Therefore, new closed geodesics all have lengths greater than $L_c$.

\begin{case}\label{case4}
$\Delta_c{\cal V}$ belongs to $\partial\cal V$.
\end{case}
Here, two $c$-borders of a single surface are twist-plumbed together to give a new surface. The proof is exactly the same as Case \ref{case3}. This completes the proof of Theorem \ref{vertextheo}.

\begin{figure}[h!]
\centering
\includegraphics[width=1\textwidth]{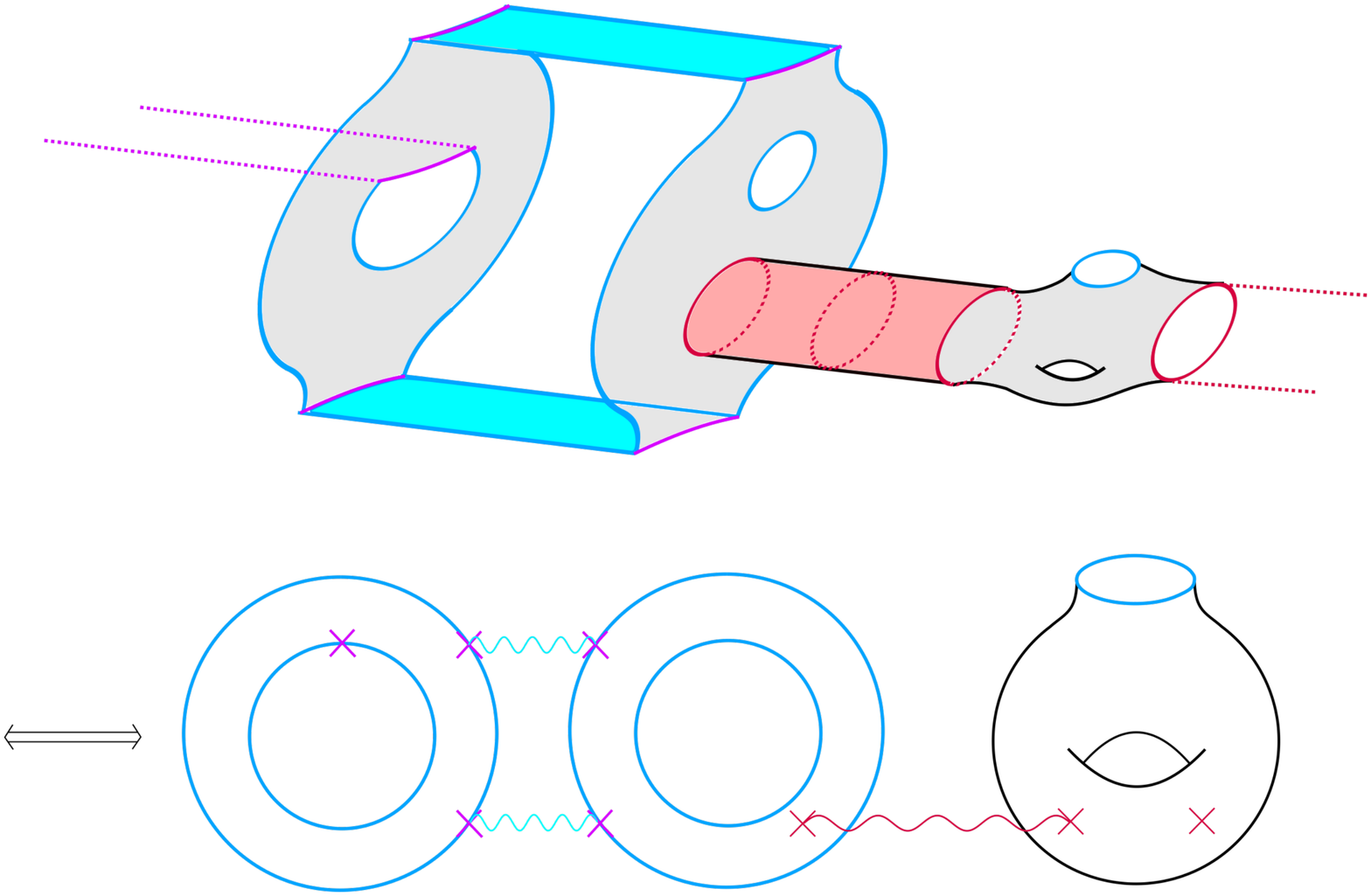}
\caption{Vertices (in gray) connected by flat finite strips corresponding to open string propagators (in skyblue) and flat finite cylinders corresponding to closed string propagators (in cherry). Open string propagators connect pairs of $p$-sides (in purple) and closed string propagators connect pairs of $c$-borders (in cherry). Grafting should also be applied (in dotted purple and cherry). $b$-sides and $b$-borders (in blue) remain as part of the boundaries. The result is the Feynman diagram on the bottom.}\label{fig:feynman}
\end{figure}

\subsection{Feynman diagrams}
Now that open-closed string vertices are explicitly constructed, we discuss how to acquire Feynman diagrams. The discussion here closely follows Section 5 of \cite{Costello:2019fuh}. There are two kinds of propagators to consider: closed string and open string. As suggested in \cite{Costello:2019fuh}, closed string propagators will correspond to flat finite cylinders of circumference $L_c$, height $t>0$, and twist angle $0\leq\theta<2\pi$. Similarly, open string propagators will correspond to flat finite strips of width $L_o$ and height $h>0$. Using these propagators, we can form Feynman diagrams by grafting cylinders to pairs $c$-borders of vertices, and strips to pairs of $p$-sides of vertices. This is illustrated in Figure \ref{fig:feynman}.

Therefore, the metric is hyperbolic over vertex regions while flat over propagator regions. Even though such a metric is not smooth over $c-$borders/$p$-sides used for plumbing/gluing as the curvature jumps, it is nonetheless continuous. It is named Thurston metric and we refer readers to \cite{tanigawa1995grafting} for a review.

Recall Theorem \ref{fundamentalhomology} that chains acquired by Feynman diagrams using solutions to open-closed geometric master equation (\ref{geometricmaster}) represent the fundamental class of the homology relative to the boundary upon push-forward to the moduli space. If it further happens that the Feynman diagrams built out of hyperbolic vertices are sections, then they provide a decomposition of the moduli space of bordered Riemann surfaces with bulk and boundary punctures. In particular, it means that the parameters of propagators and vertices become injective coordinates of the moduli space. We do not have any strong evidence for such an argument at this point, but it will be interesting to see if this is the case explicitly for some low genus, low number of boundaries, and low number of bulk and boundary punctures.

\section{Description of zero and one-dimensional vertices}\label{lowdimvertices}
In this section, we describe low dimensional open-closed hyperbolic string vertices explicitly. For the convenience of discussions and presentations, we will describe the construction of open-closed hyperbolic surfaces in $H_{L_o,L_c}$ corresponding to vertices, without mentioning graftings for the punctures, which is always obvious and assumed.

\subsection{Dimension zero vertices}
We already constructed a dimension zero vertex: disk with one closed string puncture (\ref{disk1pt}). It is important to note that this has one conformal Killing vector. Therefore, when we plumb the closed string puncture on this disk with a closed string puncture of another surface, the dimension of the moduli will increase only by one, rather than two. Other dimension zero vertices below will not have any conformal Killing vector.

\begin{figure}[h!]
\centering
\includegraphics[width=1\textwidth]{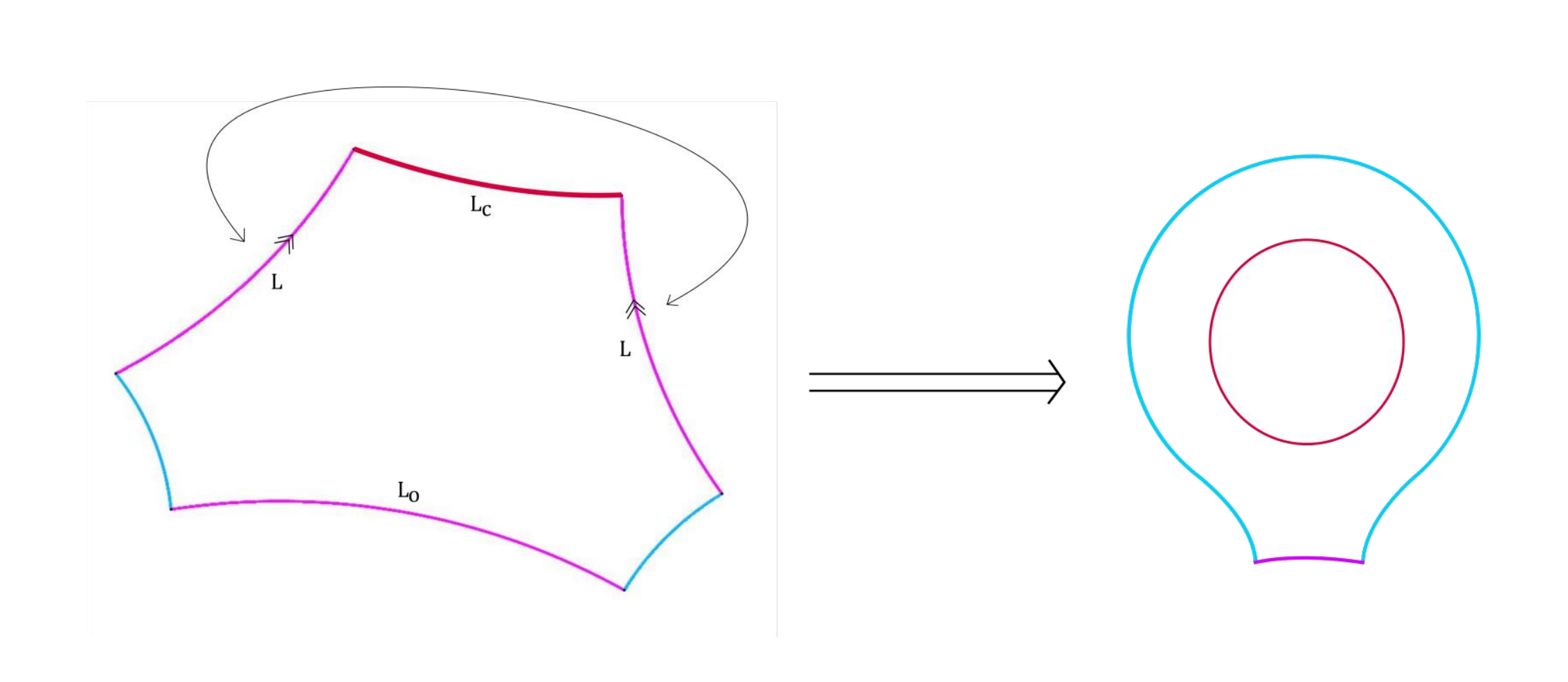}
\caption{Disk with one bulk puncture and one boundary puncture formed by a $p$-side gluing of a hexagon. Two $p$-sides (in purple) of lengths $L$ are glued together and the side between them (cherry) of length $L_c$ becomes a $c$-border. Lengths $L$ and $L_c$ satisfy the relation (\ref{diskbulkbdryeqn}).}\label{fig:diskbulkbdry}
\end{figure}

\textbf{i) Disk with three open string punctures}

This is our favorite right-angled hexagon, with all $p$-sides being of lengths $L_o$. Note that there are two marking inequivalent hexagons to consider.

\textbf{ii) Sphere with three closed string punctures}

This is a Y-piece, with three closed boundary geodesics all being $c$-borders of lengths $L_c$.

\textbf{iii) Disk with one closed string puncture and one open string puncture}

This can be built using a hexagon drawn in Figure \ref{fig:diskbulkbdry}. One of three $p$-sides has length $L_o$, corresponding to the open string puncture. The other two $p$-sides have lengths $L$, which is given by
\ie\label{diskbulkbdryeqn}
\cosh^2L={\cosh L_c+\cosh L_o\over\cosh L_c -1}.
\fe
This is such that the $b$-side neighboring the two $p$-sides has length $L_c$, as can be checked using trigonometric identity (\ref{trigid}). We then glue two $p$-sides of lengths $L$ and the $b$-side between the two becomes a $c$-border, which corresponds to a closed string puncture.

\subsection{Dimension one vertices}
There are four vertices at dimension one, all of which have no conformal Killing vector.

\begin{figure}[h!]
\centering
\includegraphics[width=0.6\textwidth]{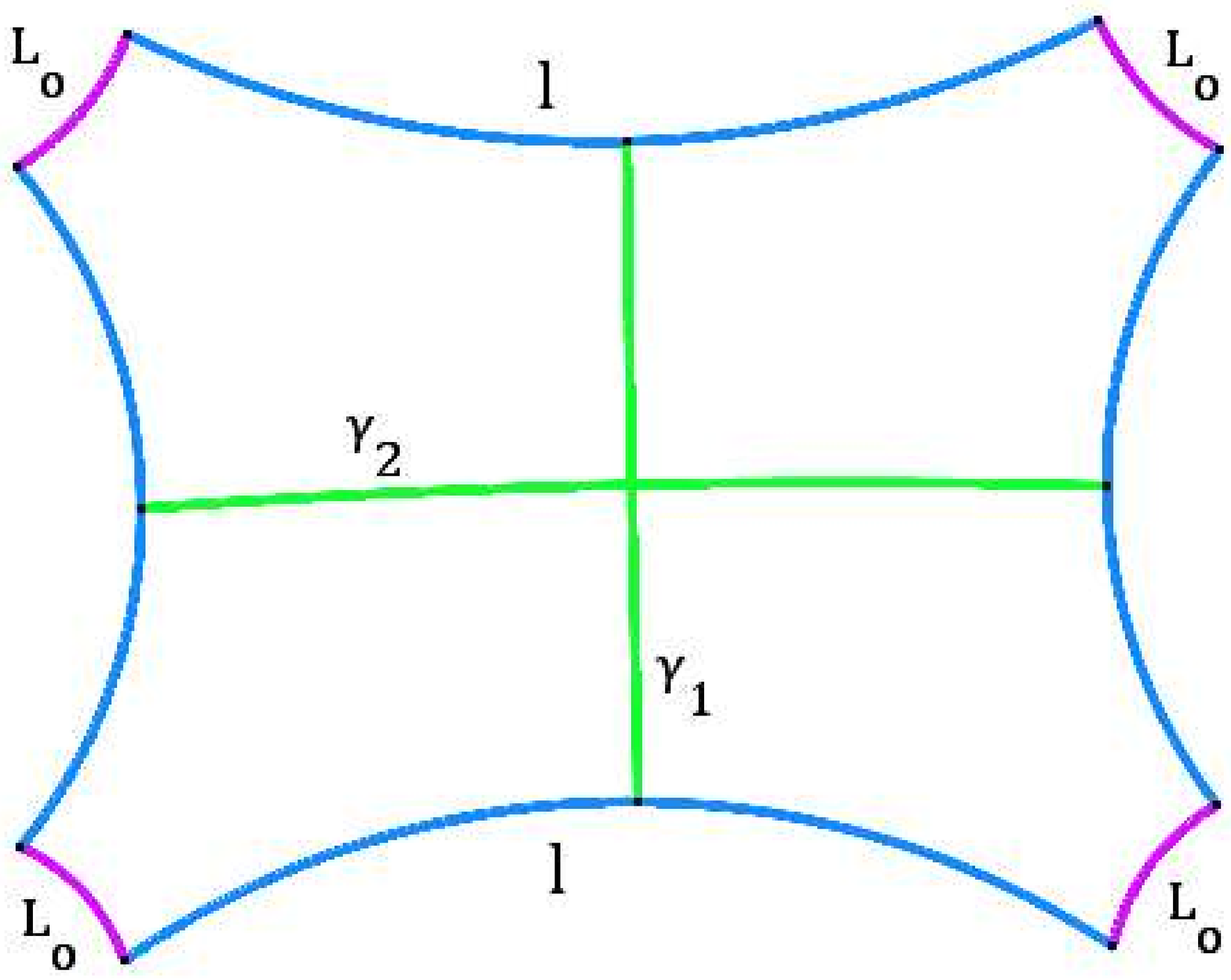}
\caption{Disk with four $p$-sides of lengths $L_o$ (in purple). There are two $p$-geodesics to consider (in green) and either of their lengths provides a coordinate for the moduli space. Lengths of the top and bottom $b$-sides are denoted as $l$.}\label{fig:open4pt}
\end{figure}

\textbf{i) Disk with four open string punctures}

For the convenience of discussion and presentation, we are going to work with only one ordering of markings, say $\{1,2,3,4\}$ for the open string punctures in the clockwise direction. The other five marking inequivalent diagrams can be obtained simply by permuting the marking labels. We take two identical hexagons whose two of the $p$-sides are of lengths $L_o$ corresponding to open string punctures, and the other $p$-side of length $L$. We then glue along the $p$-side of length $L$ as shown in Figure \ref{fig:open4pt}. This $p$-side becomes a $p$-geodesic of the glued surface, which we call $\gamma_1$. Thus, we have the first condition for the vertex: $L\geq L_o$. However, this is not the only $p$-geodesic which is not a $p$-side. There is another $p$-geodesic $\gamma_2$ of length $L'$, which crosses $\gamma_1$, and thus we have the second vertex condition $L'\geq L_o$. Lengths described in Figure \ref{fig:open4pt} have relations with each other. First, trigonometric identity for the hexagon (\ref{trigid}) implies
\ie
\cosh{l\over2}={\cosh{L_o}(1+\cosh{L})\over\sinh{L_o}\sinh{L}}.
\fe
Secondly, trigonometric identity for pentagon (\ref{pentid}) gives
\ie
\cosh{L'\over2}=\sinh{L_o}\sinh{l\over2}.
\fe
Combining the two, we get the following expression for $L'$ in terms of $L$ and $L_o$
\ie
\cosh{L'}={2\cosh(2L_o)+\cosh L+1\over\cosh L-1}.
\fe
Thus, $L$ grows monotonically as $L'$ decreases and vice versa, and they become equal at
\ie
\cosh L=\cosh L'=1+2\cosh L_o>\cosh L_o,
\fe
where we wrote the last inequality to stress that at this length, both $p$-geodesics have lengths greater than $L_o$.

Two $p$-geodesics $\gamma_1$ and $\gamma_2$ are distinct due to markings and the moduli space is parameterized either by $L$ or $L'$. We choose to work with $L$ and the full moduli space is given by $L\in\mathbb{R}_+$. With the first and second vertex conditions and the expression for $L'$ as a function of $L$, we get the following result for the vertex region
\ie\label{condition4pt}
{\cal V}^{0,0}_{1,\{4\}}\text{ condition: }L_o\leq  L\leq e(L_o),~~\text{where }\cosh\left(e(L_o)\right)\equiv{2\cosh(2L_o)+\cosh L_o+1\over\cosh L_o-1}.
\fe
Note that for any $L_o\in\mathbb{R}_+$, the vertex region for $L$ is nonempty. Again, there are five other marking inequivalent diagrams which essentially carry the exactly same form of the vertex conditions.

\begin{figure}[h!]
\centering
\includegraphics[width=1\textwidth]{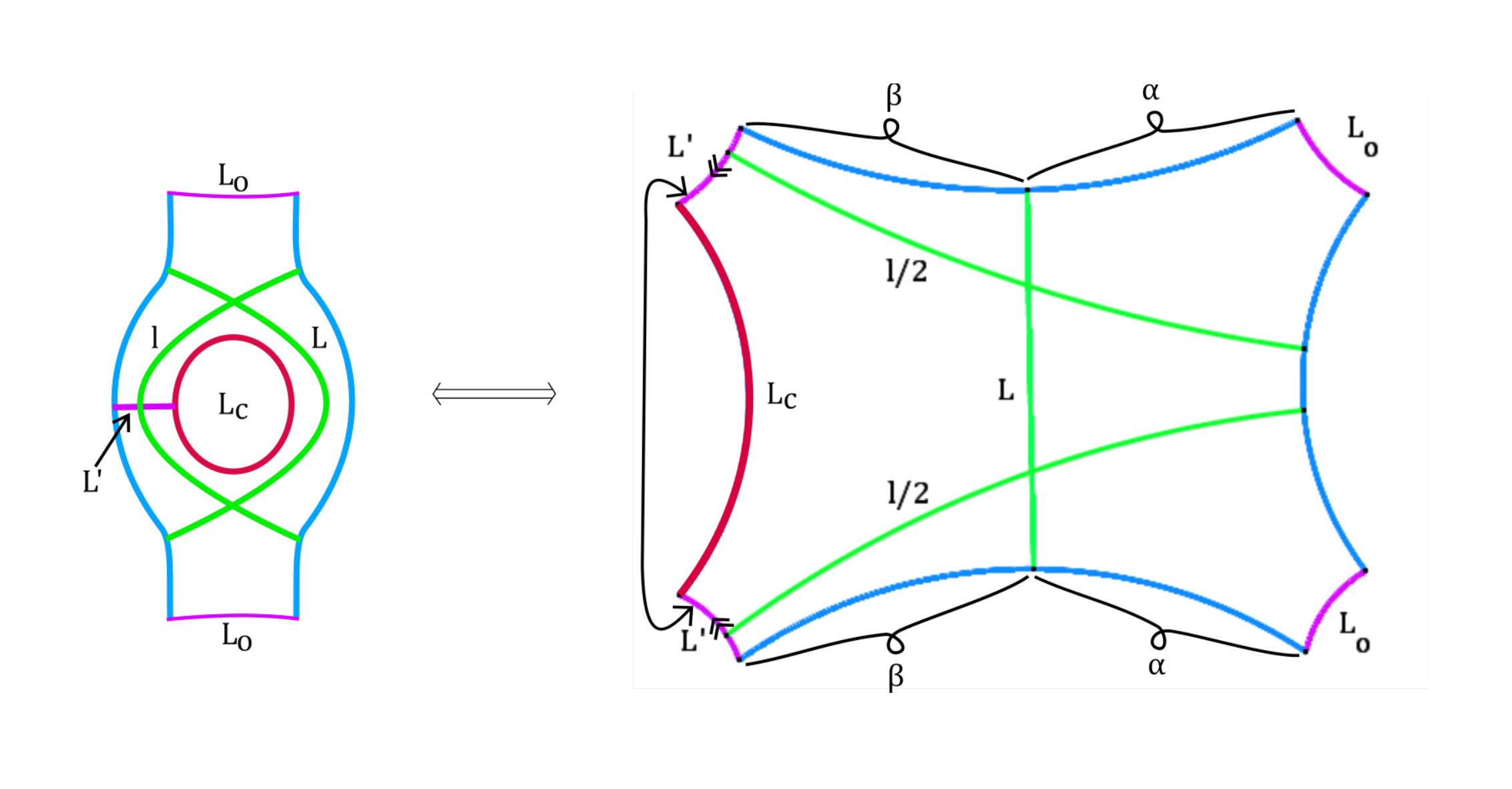}
\caption{Disk with two $p$-sides of lengths $L_o$ (in purple) and a $c$-border of length $L_c$ (in cherry). There are two $p$-geodesics of lengths $l$ and $L$ to consider (in green). Lengths $l$ and $L'$ can be expressed as functions of $L_o,L_c,$ and $L$, and the moduli space is parameterized by either $l$ or $L$.}\label{fig:diskioneclosedtwoopen}
\end{figure}

\textbf{ii) Disk with one closed string puncture and two open string punctures}

As shown in Figure \ref{fig:diskioneclosedtwoopen}, there are two $p$-geodesics to consider. We describe this case using two hexagons. The first hexagon on the left has three $p$-sides of lengths $L, L', L'$, so that two of them are equal in lengths. The $b$-side between them has length $L_c$ so that upon self-gluing two $p$-sides of lengths $L'$, we create a $c$-border of length $L_c$. This will correspond to the closed string puncture. (\ref{trigid}) gives the relation
\ie\label{relation1}
\cosh L_c={\cosh L+\cosh^2L'\over\sinh^2L'}~\Rightarrow~\cosh^2L'={\cosh L_c+\cosh L\over\cosh L_c-1}.
\fe

The second hexagon on the right has $p$-sides of lengths $L, L_o, L_o$. Two $p$-sides with lengths $L_o$ correspond to two open string punctures. We now glue two $p$-sides of lengths $L$ of two hexagons together. Then, it becomes a $p$-geodesic, whose endpoints are on two $b$-sides neighboring $p$-sides of lengths $L'$ and $L_o$, and we have the first vertex condition $L\geq L_o$. The other $p$-geodesic has length $l$ and has its endpoints on a $b$-side between two $p$-sides of $L_o$. Therefore, the second vertex condition is given by $l\geq L_o$. We now find the relation between the two lengths $L$ and $l$.

Again using (\ref{trigid}), we get
\ie
\cosh\alpha={\cosh L_o(1+\cosh L)\over\sinh L_o\sinh L},~~\cosh\beta={\cosh L'(1+\cosh L)\over\sinh L'\sinh L}={\cosh L+1\over\sinh L}\sqrt{\cosh L+\cosh L_c\over\cosh L+1},
\fe
where we used (\ref{relation1}) in the last equality. Also, trigonometric identity for pentagon (\ref{pentid}) gives the following relation
\ie
\cosh{l\over2}=\sinh L_o\sinh(\alpha+\beta).
\fe
Combining altogether, we get
\small
\ie\label{relation2}
\cosh l=-1+2\coth^2{L\over2}\left(\sqrt{1+\cosh L_c\over\cosh L-1}\cosh L_o+\sqrt{(\cosh L+\cosh L_c)(\coth^2L_o\coth^2{L\over2}-1)\over\cosh L+1}\sinh L_o \right)^2.
\fe
\normalsize
It is straighforward to see that as $L$ increases from $0$ to $\infty$, $l$ monotonically decreases from $\infty$ to $0$. The moduli space is parameterized by either $L$ or $l$ since these are distinct due to markings and the orientation. We choose to work with $L$. The full moduli space is given by $L \in\mathbb{R}_+$, and the first and second vertex conditions together with the expression of $l$ as a function of $L$ give the following vertex region
\ie\label{condition1c2o}
&{\cal V}^{0,1}_{1,\{2\}}\text{ condition: }L_o\leq L\leq f(L_o,L_c),~\text{where}
\\
&\cosh\left(f(L_o,L_c)\right)\equiv-1
\\
&~~~~~~+2\coth^2{L_o\over2}\left(\sqrt{1+\cosh L_c\over\cosh L_o-1}\cosh L_o+\sqrt{(\cosh L_o+\cosh L_c)(\coth^2L_o\coth^2{L_o\over2}-1)\over\cosh L_o+1}\sinh L_o \right)^2.
\fe
One can check that this region is nonempty for all $(L_o,L_c)\in{\cal R}$.

\begin{figure}[h!]
\centering
\includegraphics[width=1\textwidth]{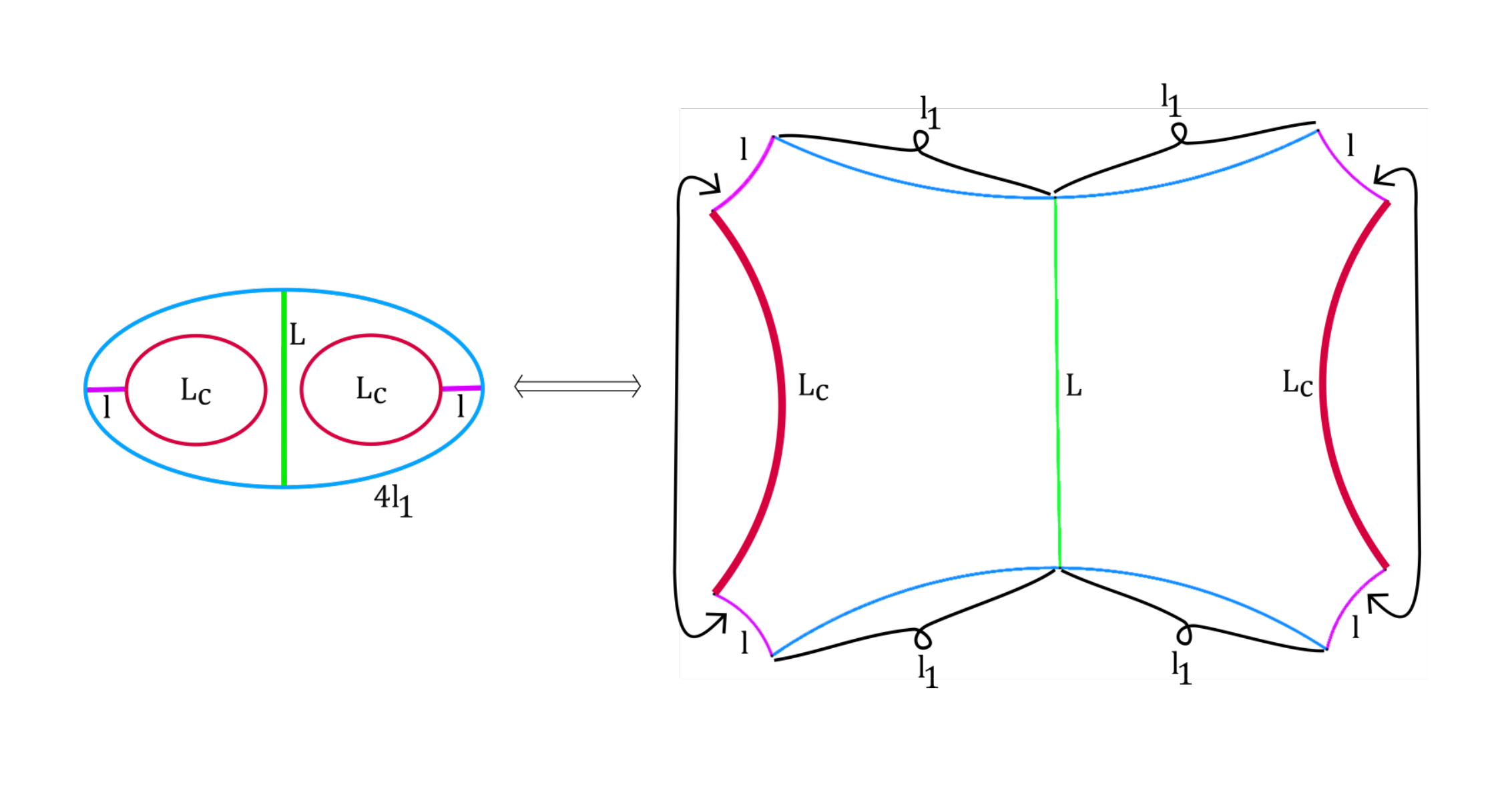}
\caption{Disk with two $c$-borders of lengths $L_c$ (in cherry). There is a single $p$-geodesic of length $L$ to consider (in green) which parameterizes the moduli space. $b$-border (in blue) has length $4l_1$ which is a function of $L$ and $L_c$.}\label{fig:disktwoclosed}
\end{figure}

\textbf{iii) Disk with two closed string punctures}

We take two identical hexagons with $p$-sides of lengths $L, l, l$ as shown in Figure \ref{fig:disktwoclosed}. The $b$-side between two $p$-sides of length $l$ will have length $L_c$, and (\ref{trigid}) implies that the other two left-over $b$-sides have equal lengths which we call $l_1$. Using (\ref{trigid}), we have the relations $\cosh^2l={\cosh{L_c}+\cosh{L}\over\cosh L_c-1}$ and $\cosh l_1={\cosh l(1+\cosh L)\over\sinh l\sinh L}$. Combining these, we get
\ie\label{l1id}
\cosh^2l_1={\cosh L+\cosh L_c\over\cosh L-1}.
\fe

For each hexagon, we glue two $p$-sides of lengths $l$ together to form a $c$-border of length $L_c$. This will correspond to a closed string puncture, so each hexagon has one closed string puncture. Now, we also glue $p$-sides of length $L$ of two hexagons together, which is the only $p$-geodesic of the resulting surface which is not a $p$-side. Therefore, the moduli space is given by $L\in\mathbb R_+$. We have the first vertex condition $L\geq L_o$. Also, we have a closed geodesic which is the boundary of the surface i.e. $b$-border. Its length is $4l_1$ so we require $4l_1\geq L_c$. Using (\ref{l1id}), we get the following vertex condition
\ie\label{condition2c}
{\cal V}^{0,2}_{1,\{0\}}\text{ condition: } L_o\leq L\leq g(L_c),~~\text{where }\cosh\left(g(L_c)\right)\equiv{\cosh^2{L_c\over4}+\cosh L_c\over\cosh^2 {L_c\over4}-1}.
\fe
Note that this is nonempty for all $(L_o,L_c)\in\cal R$.

\begin{figure}[h!]
\centering
\includegraphics[width=1\textwidth]{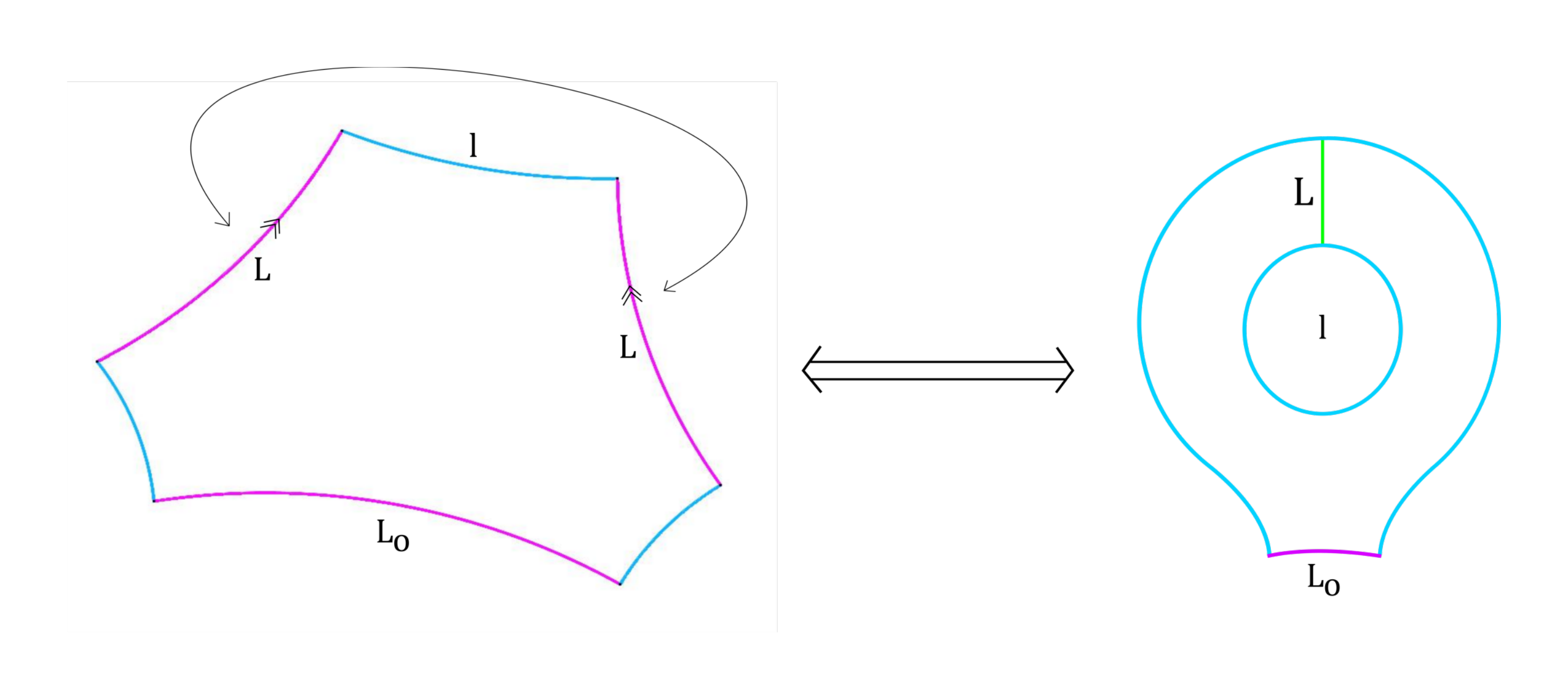}
\caption{Annulus with one $p$-side of length $L_o$ (in purple) constructed from a hexagon with two $p$-sides of lengths $L$ glued together, which then become a $p$-geodesic (in green). $b$-border of length $l$ (in blue) is also formed and the moduli space can be parameterized by either $l$ or $L$. We choose to work with $L$.}\label{fig:annulus1pt}
\end{figure}

\textbf{iv) Annulus with one open string puncture}

We take a hexagon with $p$-sides of lengths $L_o,L,L$ as shown in Figure \ref{fig:annulus1pt}. The $p$-side with length $L_o$ correspond to the open string puncture. The $b$-side between two $p$-sides of lengths $L$ has length $l$ satisfying $\cosh l={\cosh L_o+\cosh^2L\over\sinh^2L}$ due to (\ref{trigid}).

We glue two $p$-sides of lengths $L$, which then becomes the only $p$-geodesic which is not a $p$-side. Thus, the moduli space is parameterized by $L\in\mathbb R_+$. We have the first vertex condition $L\geq L_o$. Also, the $b$-side of length $l$ becomes a $b$-border, so it should satisfy $l\geq L_c$. Using the previous length relation, we get
\ie\label{conditionAnn}
{\cal V}^{0,0}_{2,\{1,0\}}\text{ condition: }L_o\leq L\leq h(L_o,L_c),~~\text{where }\cosh\left( h(L_o,L_c)\right)\equiv\sqrt{\cosh L_c+\cosh L_o\over\cosh L_c-1}.
\fe
Again, this is nonempty for all $(L_o,L_c)\in\cal R$.

This completes the construction of open-closed hyperbolic string vertices of dimension up to one.

\begin{figure}[h!]
\centering
\includegraphics[width=1.1\textwidth]{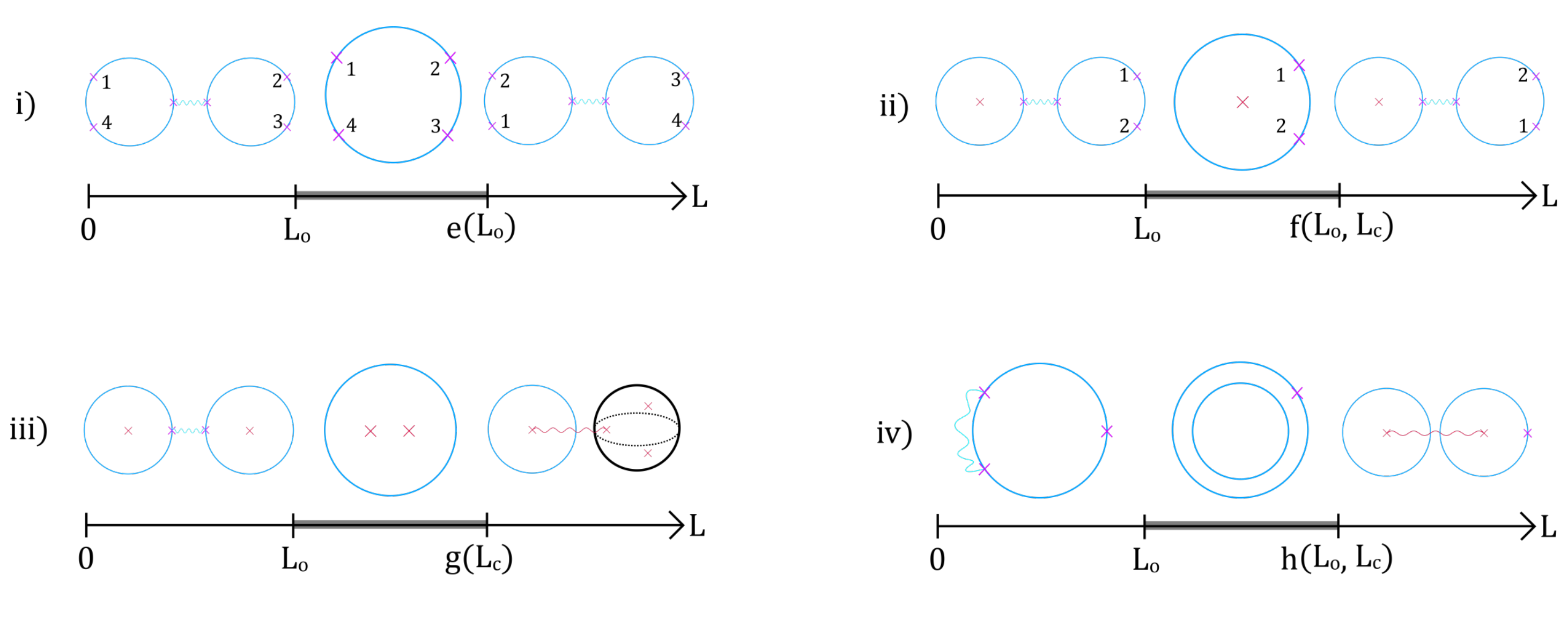}
\caption{One-dimensional moduli spaces assuming that Feynman diagrams provide sections. Different regions are covered by either vertices (colored in gray) or Feynman diagrams with propagators. Note that i) is a single piece of six marking inequivalent configurations, while the cases ii), iii), and iv) have unique markings.}\label{fig:modulispace}
\end{figure}

\subsection{Decomposition and Deligne-Mumford compactification of moduli spaces of dimension one}
In this subsection, we will assume that Feynman diagrams built by hyperbolic vertices indeed provide sections, and study which diagrams cover which parts of the moduli spaces for the case of dimension one. All four cases below are summarized in Figure \ref{fig:modulispace} and notations for each case follow those in the previous subsection. We also discuss Deligne-Mumford compactifications of the moduli spaces.

\textbf{i) Disk with four open string punctures}

The regions $L<L_o$ and $L>e(L_o)$ should be covered by Feynman diagrams with two cubic open string vertices ${\cal V}^{0,0}_{1,\{3\}}$ and an open string propagator connecting the two. Two regions carry different markings as shown in Figure \ref{fig:modulispace}. In a sense, open string propagator shortens the $p$-geodesic.

\textbf{ii) Disk with one closed string puncture and two open string punctures}

The regions $L<L_o$ and $L>f(L_o,L_c)$ are covered by Feynman diagrams with two vertices ${\cal V}^{0,1}_{1,\{1\}}$ and ${\cal V}^{0,0}_{1,\{3\}}$ with an open string propagator between them, where the markings on ${\cal V}^{0,0}_{1,\{3\}}$ are different between the two regions. Again, open string propagator effectively shortens the $p$-geodesic.

\textbf{iii) Disk with two closed string punctures}

There are two Feynman diagrams other than the vertex. The first one is given by two identical vertices ${\cal V}^{0,1}_{1,\{1\}}$ with an open string propagator. This effectively shortens the length $L$ of the $p$-geodesic and covers $L<L_o$. The second one is given by two vertices ${\cal V}^{0,1}_{1,\{0\}}$ and ${\cal V}^{0,3}_{0,\{\}}$ with a closed string propagator. Here, Y-piece with three $c$-borders of lengths $L_c$ acquires a boundary as the closed string propagator is attached to one of $c$-borders, since the other end of the propagator sticks to a circle of circumference $L_c$ and becomes a $b$-border. This effectively reduces the length of the plumbed $c$-border which is a $b$-border now, thus covering $4l_1<L_c$.

\textbf{iv) Annulus with one open string puncture}

There are again two Feynman diagrams other than the vertex. One is given by a vertex ${\cal V}^{0,0}_{1,\{3\}}$ with an open string propagator. This effectively shortens the length $L$ of the $p$-geodesic and covers the moduli space $L<L_o$. Another diagram consists of two vertices ${\cal V}^{0,1}_{1,\{0\}}$ and ${\cal V}^{0,1}_{1,\{1\}}$ with a closed string propagator, which effectively shortens the length $l$ of the $b$-border. Thus, it covers $l<L_c$.

Finally, note that we have simple descriptions of Deligne-Mumford compactifications for all four cases. In cases i) and ii), the compactification adds a point $L=0$ corresponding to the infinite open string propagator and another point $L=\infty$ also corresponding to the infinite open string propagator. In cases iii) and iv), the compactification similarly adds $L=0$ for the infinite open string propagator, but this time adds $L=\infty$ for the infinite closed string propagator. The latter is typically of codimension two and thus one might naively think that it cannot add a point to one-dimensional moduli space. However, due to the presence of one conformal Killing vector on a disk with a bulk puncture, it indeed adds a point which becomes a boundary of the moduli space. Therefore, all the one-dimensional moduli spaces $L\in(0,\infty)$ become $L\in[0,\infty]$ under Deligne-Mumford compactifications, where again, ${\cal M}^{0,0}_{1,\{4\}}$ strictly speaking should also include other five  marking inequivalent configurations. As expected, Deligne-Mumford compactifications indeed get nontrivial boundary contributions, which is not the case for closed Riemann surfaces with punctures.

\subsection{No cubic point in $\cal R$}\label{cubic}
It is well known that classical open string field theory has a representation where it has only a cubic vertex \cite{Witten:1985cc}. In the case of open-closed string vertices, analogous cubic theory was constructed using minimal area metric in \cite{Zwiebach:1992bw}, with one caveat that closed strings were always taken to be on-shell. The idea was that with a specific choice of a minimal area metric, all one-dimensional moduli spaces were covered by Feynman diagrams of zero dimensional vertices, meaning that one-dimensional vertices are empty. With the condition that closed string fields are on-shell and do not propagate in particular, it is obvious from geometric master equation that all higher-dimensional vertices whose surfaces have at least one boundary can be set to be empty, as it relates boundary of $d$-dimensional vertices to $\{~,~\}$ acting on two vertices whose dimensions add up to $(d-1)$ or $\Delta$ acting on a single $(d-1)$-dimensional vertex.

In our construction of open-closed hyperbolic vertices, we explicitly saw that all one-dimensional vertices are nonempty. For example, consider ${\cal V}^{0,0}_{1,\{4\}}$ given by (\ref{condition4pt}). For any choice of $L_o\in\mathbb R_+$, it is nonempty meaning that one cannot set the open string quartic vertex to be empty. Therefore, we conclude that our construction of hyperbolic vertices do not have a point in $\cal R$ where the theory becomes cubic.

Nonetheless, there exists a similar limit as the one described in \cite{Zwiebach:1992bw}. We can consider $L_c\rightarrow0$ where closed string propagators become infinite. In this limit, closed string fields must be on-shell and do not propagate. Even though all the vertices involving both open and closed string fields are still present, moduli spaces get only infinitesimal contributions from Feynman diagrams involving closed string propagators, meaning that it is covered mostly by vertices and open string propagators. In particular, the description of dynamics of open string fields requires diagrams involving only open string fields without any closed string fields, modulo boundary contributions corresponding to infinite closed string propagators attached to vertices involving on-shell closed strings.

The other limit $L_o\rightarrow0$ describes the opposite situation where the open string propagators become infinitely lengthy. In this case, moduli spaces get infinitesimal contributions from Feynman diagrams involving open string propagators and in particular, dynamics of closed string fields are described by diagrams involving only closed string fields, modulo boundary contributions from infinite open string propagators attached to vertices involving on-shell open strings. Of course, it is important to keep in mind that these limits are only formal and the full open-closed string field theory requires finite nonzero $L_c$ and $L_o$.

\section{Discussions}\label{discussions}
In this work, we found a family of hyperbolic string vertices for open-closed string field theory, naturally generalizing the work of \cite{Costello:2019fuh} which considered the pure closed string vertices. Key ingredients were collar theorems of BHHS, which are responsible for restricting the systolic conditions on nontrivial simple closed geodesics and $p$-geodesics to region $\cal R$ (\ref{regionR}). We then gave explicit descriptions of all zero and one-dimensional vertices.

There are two mathematical steps which one could naturally try to take. First is the existence and uniqueness of the solutions to the geometric master equation (\ref{geometricmaster}), following similar ideas as \cite{Costello:2019fuh}. There may be subtleties due to the fact that Deligne-Mumford compactifications get nontrivial boundary contributions, but the general idea seems to allow for a rather straighfoward generalization to the open-closed case. The second is to check if the grafting map provides a homeomorphism between ${\cal{M}}_{b,\{m_i\}}^{g,n}(L_o,L_c)$ and ${\cal{M}}_{b,\{m_i\}}^{g,n}$, and if hyperbolic Feynman diagrams provide sections, which will then lead to coordinates for the entire moduli space.

On the computational side, in order to actually compute vertices with string field insertions, one needs to learn how to compute conformal field theory correlation functions on grafted BHHS, which will involve nontrivial Weyl transformations due to the presence of geodesic sides and borders. In addition, one then has to integrate such correlation functions over the vertex regions in moduli spaces. Both of these technical aspects are less explored in literature and thus remain as main obstacles if we were to use hyperbolic string vertices to perform string field theory computations.

In the case of moduli integration for hyperbolic surfaces, the results are known only for simple cases such as volumes over bordered hyperbolic Riemann surfaces via topological recursion relations \cite{1998InMat.132..607M,Mirzakhani:2006fta,Mirzakhani:2006eta}. Recall that the Weil-Petersson volume form for bordered hyperbolic Riemann surfaces is built from wedge products of a symplectic form expressed in terms of Fenchel-Nielsen coordinates, which is independent of different possible pants decompositions. In the case of BHHS however, it does not seem so straightforward to find a simple expression for the volume form which should be independent of the hexagon decompositions. For example, such a form for ${\cal M}^{0,0}_{1,\{4\}}$ is given by ${d(\cosh L)\over\cosh L-1}$, but if the lengths of the $p$-sides are different from each other, this volume form is no longer invariant under different hexagon decompositions. Nonetheless, hexagon decompositions for BHHS seem to provide another setup to study the moduli spaces of bordered hyperbolic Riemann surfaces in a rather direct way. Perhaps, it is not a coincidence that ${\cal N}=4$ super Yang-Mills integrability story also benefited by considering hexagons \cite{Basso:2015zoa,Basso:2015eqa,Fleury:2016ykk,Fleury:2017eph}. Interestingly, the four-bordered sphere of Figure 1 in \cite{Fleury:2016ykk} is exactly $p$-side gluing of four hexagons, rather than typical plumbing construction of two Y-pieces.

On the physics side, open-closed string field theory has several interesting applications. Even at the level of computing perturbative on-shell ampltiudes involving D-branes \cite{Hashimoto:1996bf} or D-instanton contributions to type IIB closed string scattering amplitudes \cite{Polchinski:1994fq,Green:1997tv}, unambiguous procedures are presumably defined only through open-closed string field theory. Quantum master action for open-closed superstring field theory for instance was constructed only recently in \cite{Moosavian:2019ydz} and consistent computations of on-shell amplitudes using such a framework are yet to be performed. It will be interesting to explicitly see how open-closed string field theory resolves issues related to boundary contributions of the moduli space discussed in section \ref{fundclass}. This issue has been recently discussed in the context of ZZ instanton contributions to $c=1$ string theory \cite{Balthazar:2019rnh} using the open-closed string field theory framework \cite{Sen:2019qqg}.

Descriptions of open string tachyon dynamics \cite{Sen:2002nu,Sen:2002in,Sen:2003xs,Sen:2003iv,Sen:2004nf}, S-brane solutions \cite{Gutperle:2002ai}, and open-closed type dualities \cite{Kontsevich:1992ti,Witten:1992fb,Gopakumar:1998ki,Khoury:2000hz,Ooguri:2002gx,Gaiotto:2001ji,Gaiotto:2003rm,Gaiotto_2005} may be formulated in the framework open-closed string field theory. For example, unstable D-brane decay is classically described by an appropriate time-dependent boundary state and it was conjectured \cite{Sen:2003iv} that the full physics describing the unstable D-brane is given by a quantum open string field theory, not necessarily needing to introduce closed strings into the description. It does not seem unrelated to the limit $L_c\rightarrow0$ discussed in section \ref{cubic}, even though the limit was only a formal one and the precise open string field theory is yet to be understood. On the open-closed duality side, one of typical ideas is that on-shell closed string processes can be constructed from open string diagrams. For example, one can take a vacuum open string amplitude and let the boundaries shrink to become on-shell closed string punctures. This corresponds to plumbing disks with one closed string puncture to pure closed string diagrams via infinite closed string propagators. Along the similar line, one can also take a solution to open-closed string field equations of motions and study how it interpolates between different looking solutions under field redefinitions induced by shifts in vertices e.g. taking different values of $L_o$ and $L_c$ in region $\cal R$ of our hyperbolic vertices. In particular, limits $L_c\rightarrow0$ and $L_o\rightarrow0$ would respectively give mostly open and mostly closed solution in the sense that $L_c\rightarrow0$ turns off off-shell closed string field deformation and $L_o\rightarrow0$ turns off off-shell open string field deformation from the original boundary conformal field theory around which open-closed string field theory was formulated\footnote{We thank Xi Yin for suggesting viewing dual descriptions of open string and closed string field solutions as string field redefinitions interpolating between mostly open and mostly closed pictures.}. In many respects, it definitely appears that string vertices and string field theory in general provide a rich list of relevant mathematical problems and interesting physics questions to be answered.

\section*{Acknowledgments}
We would like to thank Ashoke Sen, Arnav Tripathy, Xi Yin, Max Zimet, and Barton Zwiebach for enlightening discussions and correspondences, and Ashoke Sen, Xi Yin, and Barton Zwiebach for comments on an early draft. This work is supported in part by Xi Yin's Simons Investigator Award from the Simons Foundation and the Simons Collaboration Grant on the Non-Perturbative Bootstrap.

\appendix
\section{Hyperbolic surfaces}\label{hyp}
Here, we collect relevant theorems and identities appearing in hyperbolic surfaces. All of these statements are explained in detail in \cite{Buser1992GeometryAS}. For the following theorems, we assume that we are on a general compact hyperbolic surface $S$.
\begin{theo}\label{perpendiculartheorem}{\normalfont{(Theorem 1.5.3 in \cite{Buser1992GeometryAS})}}
Given $A$ and $B$ which are either smooth closed boundary geodesics or geodesic sides which meet neighboring sides under an angle $\leq\pi/2$, and an open curve $c$ ending on them, there exists a unique shortest geodesic $\gamma$ in the homotopy class of $c$, and $\gamma$ is the unique common perpendicular to $A$ and $B$ in the homotopy class unless $\gamma$ is a point. In particular, if $c$ were simple, $\gamma$ is also simple.
\end{theo}
This theorem is the reason why we favored right-angled hexagons and hexagon decompositions as they naturally consist of such perpendicular geodesics. The following theorem further justifies some of considerations of closed geodesics in our construction of hyperbolic vertices.
\begin{theo}\label{closedgeotehroem}{\normalfont (Theorem 1.6.6 in \cite{Buser1992GeometryAS})}
Given a nontrivial closed curve $c$ on $S$, it is freely homotopic to a unique closed geodesic $\gamma$. $\gamma$ is either contained in $\partial S$ or $\gamma\cap\partial S=\varnothing$.
\end{theo}

We now state some of relevant trigonometric identities in hyperbolic geometry. There are more identities than the ones we discuss here, which can all be found in Chapter 2 of \cite{Buser1992GeometryAS}. First, consider a right-angled triangle of geodesic sides of lengths $a, b, c$. Assume that $a$ and $b$ are meeting at the right angle. Then,
\ie\label{triangleid}
\cosh c=\cosh a \cosh b.
\fe
This identity states that the side away from the right angle is lengthier than the other two sides.

We also consider a right-angled pentagon consisting of five geodesic sides of lengths $a,b,c,d,e$ where we took into account the ordering e.g. $b$ meets $a$ to the left and $c$ to the right. All sides meet their neighboring sides at right angles. Then,
\ie\label{pentid}
\cosh c=\sinh a\sinh e,
\fe
and likewise for the other sides.

\bibliographystyle{JHEP}
\bibliography{open_closed}

\providecommand{\href}[2]{#2}\begingroup\raggedright\begin{thebibliography}{10}

\bibitem{Zwiebach:1992ie}
B.~Zwiebach, {\it {Closed string field theory: Quantum action and the B-V
  master equation}},  {\em Nucl. Phys.} {\bf B390} (1993) 33--152,
  [\href{http://arxiv.org/abs/hep-th/9206084}{{\tt hep-th/9206084}}].

\bibitem{Sen:2015uaa}
A.~Sen, {\it {BV Master Action for Heterotic and Type II String Field
  Theories}},  {\em JHEP} {\bf 02} (2016) 087,
  [\href{http://arxiv.org/abs/1508.05387}{{\tt arXiv:1508.05387}}].

\bibitem{deLacroix:2017lif}
C.~de~Lacroix, H.~Erbin, S.~P. Kashyap, A.~Sen, and M.~Verma, {\it {Closed
  Superstring Field Theory and its Applications}},  {\em Int. J. Mod. Phys.}
  {\bf A32} (2017), no.~28n29 1730021,
  [\href{http://arxiv.org/abs/1703.06410}{{\tt arXiv:1703.06410}}].

\bibitem{Sen:2019jpm}
A.~Sen, {\it {String Field Theory as World-sheet UV Regulator}},  {\em JHEP}
  {\bf 10} (2019) 119, [\href{http://arxiv.org/abs/1902.00263}{{\tt
  arXiv:1902.00263}}].

\bibitem{Cho:2018nfn}
M.~Cho, S.~Collier, and X.~Yin, {\it {Strings in Ramond-Ramond Backgrounds from
  the Neveu-Schwarz-Ramond Formalism}},
  \href{http://arxiv.org/abs/1811.00032}{{\tt arXiv:1811.00032}}.

\bibitem{Hata:1993gf}
H.~Hata and B.~Zwiebach, {\it {Developing the covariant Batalin-Vilkovisky
  approach to string theory}},  {\em Annals Phys.} {\bf 229} (1994) 177--216,
  [\href{http://arxiv.org/abs/hep-th/9301097}{{\tt hep-th/9301097}}].

\bibitem{Sen:1993ic}
A.~Sen and B.~Zwiebach, {\it {A Note on gauge transformations in
  Batalin-Vilkovisky theory}},  {\em Phys. Lett.} {\bf B320} (1994) 29--35,
  [\href{http://arxiv.org/abs/hep-th/9309027}{{\tt hep-th/9309027}}].

\bibitem{Sen:2014dqa}
A.~Sen, {\it {Gauge Invariant 1PI Effective Action for Superstring Field
  Theory}},  {\em JHEP} {\bf 06} (2015) 022,
  [\href{http://arxiv.org/abs/1411.7478}{{\tt arXiv:1411.7478}}].

\bibitem{Sen:2015hha}
A.~Sen, {\it {Gauge Invariant 1PI Effective Superstring Field Theory: Inclusion
  of the Ramond Sector}},  {\em JHEP} {\bf 08} (2015) 025,
  [\href{http://arxiv.org/abs/1501.00988}{{\tt arXiv:1501.00988}}].

\bibitem{Sen:1994kx}
A.~Sen and B.~Zwiebach, {\it {Background independent algebraic structures in
  closed string field theory}},  {\em Commun. Math. Phys.} {\bf 177} (1996)
  305--326, [\href{http://arxiv.org/abs/hep-th/9408053}{{\tt hep-th/9408053}}].

\bibitem{Sen:1993kb}
A.~Sen and B.~Zwiebach, {\it {Quantum background independence of closed string
  field theory}},  {\em Nucl. Phys.} {\bf B423} (1994) 580--630,
  [\href{http://arxiv.org/abs/hep-th/9311009}{{\tt hep-th/9311009}}].

\bibitem{Zwiebach:1990nh}
B.~Zwiebach, {\it {How covariant closed string theory solves a minimal area
  problem}},  {\em Commun. Math. Phys.} {\bf 136} (1991) 83--118.

\bibitem{Zwiebach:1990qj}
B.~Zwiebach, {\it {Quantum open string theory with manifest closed string
  factorization}},  {\em Phys. Lett.} {\bf B256} (1991) 22--29.

\bibitem{Zwiebach:1992bw}
B.~Zwiebach, {\it {Interpolating string field theories}},  {\em Mod. Phys.
  Lett.} {\bf A7} (1992) 1079--1090,
  [\href{http://arxiv.org/abs/hep-th/9202015}{{\tt hep-th/9202015}}].

\bibitem{Headrick:2018ncs}
M.~Headrick and B.~Zwiebach, {\it {Convex programs for minimal-area problems}},
   \href{http://arxiv.org/abs/1806.00449}{{\tt arXiv:1806.00449}}.

\bibitem{Headrick:2018dlw}
M.~Headrick and B.~Zwiebach, {\it {Minimal-area metrics on the Swiss cross and
  punctured torus}},  \href{http://arxiv.org/abs/1806.00450}{{\tt
  arXiv:1806.00450}}.

\bibitem{Moosavian:2017fta}
S.~F. Moosavian and R.~Pius, {\it {Hyperbolic Geometry of Superstring
  Perturbation Theory}},  \href{http://arxiv.org/abs/1703.10563}{{\tt
  arXiv:1703.10563}}.

\bibitem{Moosavian:2017qsp}
S.~F. Moosavian and R.~Pius, {\it {Hyperbolic geometry and closed bosonic
  string field theory. Part I. The string vertices via hyperbolic Riemann
  surfaces}},  {\em JHEP} {\bf 08} (2019) 157,
  [\href{http://arxiv.org/abs/1706.07366}{{\tt arXiv:1706.07366}}].

\bibitem{Moosavian:2017sev}
S.~F. Moosavian and R.~Pius, {\it {Hyperbolic geometry and closed bosonic
  string field theory. Part II. The rules for evaluating the quantum BV master
  action}},  {\em JHEP} {\bf 08} (2019) 177,
  [\href{http://arxiv.org/abs/1708.04977}{{\tt arXiv:1708.04977}}].

\bibitem{Pius:2018pqr}
R.~Pius, {\it {Quantum Closed Superstring Field Theory and Hyperbolic Geometry
  I: Construction of String Vertices}},
  \href{http://arxiv.org/abs/1808.09441}{{\tt arXiv:1808.09441}}.

\bibitem{Costello:2019fuh}
K.~Costello and B.~Zwiebach, {\it {Hyperbolic String Vertices}},
  \href{http://arxiv.org/abs/1909.00033}{{\tt arXiv:1909.00033}}.

\bibitem{tanigawa1995grafting}
H.~Tanigawa, {\it Grafting, harmonic maps, and projective structures on
  surfaces},  1995.

\bibitem{1998InMat.132..607M}
G.~{McShane}, {\it {Simple geodesics and a series constant over Teichmuller
  space}},  {\em Inventiones Mathematicae} {\bf 132} (May, 1998) 607--632.

\bibitem{Mirzakhani:2006fta}
M.~Mirzakhani, {\it {Simple geodesics and Weil-Petersson volumes of moduli
  spaces of bordered Riemann surfaces}},  {\em Invent. Math.} {\bf 167} (2006),
  no.~1 179--222.

\bibitem{Mirzakhani:2006eta}
M.~Mirzakhani, {\it {Weil-Petersson volumes and intersection theory on the
  moduli space of curves}},  {\em J. Am. Math. Soc.} {\bf 20} (2007), no.~01
  1--24.

\bibitem{Zwiebach:1997fe}
B.~Zwiebach, {\it {Oriented open - closed string theory revisited}},  {\em
  Annals Phys.} {\bf 267} (1998) 193--248,
  [\href{http://arxiv.org/abs/hep-th/9705241}{{\tt hep-th/9705241}}].

\bibitem{Buser1992GeometryAS}
P.~Buser, {\it Geometry and spectra of compact riemann surfaces},  1992.

\bibitem{Witten:1985cc}
E.~Witten, {\it {Noncommutative Geometry and String Field Theory}},  {\em Nucl.
  Phys.} {\bf B268} (1986) 253--294.

\bibitem{Mumford1983TowardsAE}
D.~Mumford, {\it Towards an enumerative geometry of the moduli space of
  curves},  1983.

\bibitem{Moosavian:2019ydz}
S.~Faroogh~Moosavian, A.~Sen, and M.~Verma, {\it {Superstring Field Theory with
  Open and Closed Strings}},  \href{http://arxiv.org/abs/1907.10632}{{\tt
  arXiv:1907.10632}}.

\bibitem{mondello2008riemann}
G.~Mondello, {\it Riemann surfaces with boundary and natural triangulations of
  the teichmueller space},  2008.

\bibitem{scannell1998grafting}
K.~P. Scannell and M.~Wolf, {\it The grafting map of teichmuller space},  1998.

\bibitem{Basso:2015zoa}
B.~Basso, S.~Komatsu, and P.~Vieira, {\it {Structure Constants and Integrable
  Bootstrap in Planar N=4 SYM Theory}},
  \href{http://arxiv.org/abs/1505.06745}{{\tt arXiv:1505.06745}}.

\bibitem{Basso:2015eqa}
B.~Basso, V.~Goncalves, S.~Komatsu, and P.~Vieira, {\it {Gluing Hexagons at
  Three Loops}},  {\em Nucl. Phys.} {\bf B907} (2016) 695--716,
  [\href{http://arxiv.org/abs/1510.01683}{{\tt arXiv:1510.01683}}].

\bibitem{Fleury:2016ykk}
T.~Fleury and S.~Komatsu, {\it {Hexagonalization of Correlation Functions}},
  {\em JHEP} {\bf 01} (2017) 130, [\href{http://arxiv.org/abs/1611.05577}{{\tt
  arXiv:1611.05577}}].

\bibitem{Fleury:2017eph}
T.~Fleury and S.~Komatsu, {\it {Hexagonalization of Correlation Functions II:
  Two-Particle Contributions}},  {\em JHEP} {\bf 02} (2018) 177,
  [\href{http://arxiv.org/abs/1711.05327}{{\tt arXiv:1711.05327}}].

\bibitem{Hashimoto:1996bf}
A.~Hashimoto and I.~R. Klebanov, {\it {Scattering of strings from D-branes}},
  {\em Nucl. Phys. Proc. Suppl.} {\bf 55} (1997) 118--133,
  [\href{http://arxiv.org/abs/hep-th/9611214}{{\tt hep-th/9611214}}].
  [,118(1996)].

\bibitem{Polchinski:1994fq}
J.~Polchinski, {\it {Combinatorics of boundaries in string theory}},  {\em
  Phys. Rev.} {\bf D50} (1994) R6041--R6045,
  [\href{http://arxiv.org/abs/hep-th/9407031}{{\tt hep-th/9407031}}].

\bibitem{Green:1997tv}
M.~B. Green and M.~Gutperle, {\it {Effects of D instantons}},  {\em Nucl.
  Phys.} {\bf B498} (1997) 195--227,
  [\href{http://arxiv.org/abs/hep-th/9701093}{{\tt hep-th/9701093}}].

\bibitem{Balthazar:2019rnh}
B.~Balthazar, V.~A. Rodriguez, and X.~Yin, {\it {ZZ Instantons and the
  Non-Perturbative Dual of $c =$ 1 String Theory}},
  \href{http://arxiv.org/abs/1907.07688}{{\tt arXiv:1907.07688}}.

\bibitem{Sen:2019qqg}
A.~Sen, {\it {Fixing an Ambiguity in Two Dimensional String Theory Using String
  Field Theory}},  \href{http://arxiv.org/abs/1908.02782}{{\tt
  arXiv:1908.02782}}.

\bibitem{Sen:2002nu}
A.~Sen, {\it {Rolling tachyon}},  {\em JHEP} {\bf 04} (2002) 048,
  [\href{http://arxiv.org/abs/hep-th/0203211}{{\tt hep-th/0203211}}].

\bibitem{Sen:2002in}
A.~Sen, {\it {Tachyon matter}},  {\em JHEP} {\bf 07} (2002) 065,
  [\href{http://arxiv.org/abs/hep-th/0203265}{{\tt hep-th/0203265}}].

\bibitem{Sen:2003xs}
A.~Sen, {\it {Open closed duality at tree level}},  {\em Phys. Rev. Lett.} {\bf
  91} (2003) 181601, [\href{http://arxiv.org/abs/hep-th/0306137}{{\tt
  hep-th/0306137}}].

\bibitem{Sen:2003iv}
A.~Sen, {\it {Open closed duality: Lessons from matrix model}},  {\em Mod.
  Phys. Lett.} {\bf A19} (2004) 841--854,
  [\href{http://arxiv.org/abs/hep-th/0308068}{{\tt hep-th/0308068}}].

\bibitem{Sen:2004nf}
A.~Sen, {\it {Tachyon dynamics in open string theory}},  {\em Int. J. Mod.
  Phys.} {\bf A20} (2005) 5513--5656,
  [\href{http://arxiv.org/abs/hep-th/0410103}{{\tt hep-th/0410103}}].
  [,207(2004)].

\bibitem{Gutperle:2002ai}
M.~Gutperle and A.~Strominger, {\it {Space - like branes}},  {\em JHEP} {\bf
  04} (2002) 018, [\href{http://arxiv.org/abs/hep-th/0202210}{{\tt
  hep-th/0202210}}].

\bibitem{Kontsevich:1992ti}
M.~Kontsevich, {\it {Intersection theory on the moduli space of curves and the
  matrix Airy function}},  {\em Commun. Math. Phys.} {\bf 147} (1992) 1--23.

\bibitem{Witten:1992fb}
E.~Witten, {\it {Chern-Simons gauge theory as a string theory}},  {\em Prog.
  Math.} {\bf 133} (1995) 637--678,
  [\href{http://arxiv.org/abs/hep-th/9207094}{{\tt hep-th/9207094}}].

\bibitem{Gopakumar:1998ki}
R.~Gopakumar and C.~Vafa, {\it {On the gauge theory / geometry
  correspondence}},  {\em Adv. Theor. Math. Phys.} {\bf 3} (1999) 1415--1443,
  [\href{http://arxiv.org/abs/hep-th/9811131}{{\tt hep-th/9811131}}]. [AMS/IP
  Stud. Adv. Math.23,45(2001)].

\bibitem{Khoury:2000hz}
J.~Khoury and H.~L. Verlinde, {\it {On open - closed string duality}},  {\em
  Adv. Theor. Math. Phys.} {\bf 3} (1999) 1893--1908,
  [\href{http://arxiv.org/abs/hep-th/0001056}{{\tt hep-th/0001056}}].

\bibitem{Ooguri:2002gx}
H.~Ooguri and C.~Vafa, {\it {World sheet derivation of a large N duality}},
  {\em Nucl. Phys.} {\bf B641} (2002) 3--34,
  [\href{http://arxiv.org/abs/hep-th/0205297}{{\tt hep-th/0205297}}].

\bibitem{Gaiotto:2001ji}
D.~Gaiotto, L.~Rastelli, A.~Sen, and B.~Zwiebach, {\it {Ghost structure and
  closed strings in vacuum string field theory}},  {\em Adv. Theor. Math.
  Phys.} {\bf 6} (2003) 403--456,
  [\href{http://arxiv.org/abs/hep-th/0111129}{{\tt hep-th/0111129}}].

\bibitem{Gaiotto:2003rm}
D.~Gaiotto, N.~Itzhaki, and L.~Rastelli, {\it {Closed strings as imaginary
  D-branes}},  {\em Nucl. Phys.} {\bf B688} (2004) 70--100,
  [\href{http://arxiv.org/abs/hep-th/0304192}{{\tt hep-th/0304192}}].

\bibitem{Gaiotto_2005}
D.~Gaiotto and L.~Rastelli, {\it A paradigm of open/closed duality liouville
  d-branes and the kontsevich model},  {\em Journal of High Energy Physics}
  {\bf 2005} (Jul, 2005) 053–053.

\end{thebibliography}\endgroup

\end{document}